\documentclass[superscriptaddress,reprint,longbibliography,citeautoscript,prx]{revtex4-1}

\usepackage{amsmath}
\usepackage[margin=0.7in]{geometry}
\usepackage{hyperref}
\usepackage{multirow}
\usepackage{newtxtext}
\usepackage{gensymb}
\usepackage{wrapfig}
\usepackage{graphicx}
\usepackage{epstopdf}
\usepackage{lipsum}
\usepackage{float}
\usepackage{subfiles}
\usepackage{cleveref}

\usepackage[textwidth=1.5cm,textsize=footnotesize]{todonotes}

\newcommand{\ECEPrinceton}{Department of Electrical and Computer Engineering, Princeton University, Princeton, New Jersey 08540, USA}

\newcommand{\ChemPrinceton}{Department of Chemistry, Princeton University, Princeton, New Jersey 08540, USA}
\newcommand{\IAC}{Princeton Materials Institute, Princeton University, Princeton, New Jersey 08540, USA}

\begin{document}
\title{Quantifying surface losses in superconducting aluminum microwave resonators}

\author{Elizabeth Hedrick}\affiliation{\ECEPrinceton}
\author{Faranak Bahrami}\affiliation{\ECEPrinceton}
\author{Alexander C. Pakpour-Tabrizi}\affiliation{\ECEPrinceton}
\author{Atharv Joshi}\affiliation{\ECEPrinceton}
\author{Q. Rumman Rahman}\affiliation{\ECEPrinceton}
\author{Ambrose Yang}\affiliation{\ECEPrinceton}
\author{Ray D. Chang}\affiliation{\ECEPrinceton}
\author{Matthew P. Bland}\affiliation{\ECEPrinceton}
\author{Apoorv Jindal}\affiliation{\ECEPrinceton}
\author{Guangming Cheng}\affiliation{\IAC}
\author{Nan Yao}\affiliation{\IAC}
\author{Robert J. Cava}\affiliation{\ChemPrinceton}
\author{Andrew A. Houck}\affiliation{\ECEPrinceton}
\author{Nathalie P. de Leon}\affiliation{\ECEPrinceton}\email{npdeleon@princeton.edu}

\begin{abstract}

The recent realization of millisecond-scale coherence with tantalum-on-silicon transmon qubits showed that depositing the Al/$\mathrm{AlO_x}$/Al Josephson junction in a high purity, ultrahigh vacuum environment was critical for achieving lifetime-limited coherence, motivating careful examination of the aluminum surface two-level system (TLS) bath. Here, we measure the microwave absorption arising from surface TLSs in superconducting aluminum resonators, following methodology developed for tantalum resonators. We vary film and surface properties and correlate microwave measurements with materials characterization. We find that the lifetimes of superconducting aluminum resonators are primarily limited by surface losses associated with TLSs in the 2.7 nm-thick native $\mathrm{AlO_x}$. Treatment with 49\% HF removes surface $\mathrm{AlO_x}$ completely; however, rapid oxide regrowth limits improvements in surface loss and long term device stability. Using these measurements we estimate that TLSs in aluminum interfaces contribute around 27\% of the relaxation rate of state-of-the-art tantalum-on-silicon qubits that incorporate aluminum-based Josephson junctions.

\end{abstract}

\maketitle



\section{Introduction}
Superconducting circuits are a leading platform for the realization of quantum processors and have achieved several landmark demonstrations, including beyond-threshold quantum error correction \cite{zhao2022realization, google2023suppressing,google2025quantum,lacroix2025scaling,ofek2016extending,sivak2023real}, long-lived quantum memories \cite{ganjam2024surpassing, krayzman2024superconducting, matanin2023toward}, and quantum simulation \cite{barends2015digital, kumar2025digital, marcos2013superconducting, kandala2017hardware}. Despite substantial progress, state-of-the-art superconducting qubits continue to suffer from dielectric loss arising from amorphous materials, impurities, and defects~\cite{biznarova2024mitigation, chang2025eliminating, crowley2023disentangling, verjauw2022path, zemlicka2023compact, zhang2024acceptor}. Recent advances in materials engineering have enabled millisecond-scale lifetimes (T$_{1}$) and coherence times (T$_{2}$) by utilizing tantalum and high purity silicon~\cite{blandbahrami20252D}, driven by precise quantification of the different contributions to loss and addressing the TLS-related loss associated with surface contamination, surface oxides, and the bulk substrate \cite{crowley2023disentangling, place2021new, wang2022towards, blandbahrami20252D, deleon2025howto}. These millisecond-scale tantalum qubits still incorporate the traditional aluminum / aluminum oxide material stack for the Josephson junction. Therefore, a natural next step is to directly quantify the TLS losses in the native aluminum oxide to determine their contributions to the performance of state-of-the-art qubits.

\begin{figure*}
    \centering
    \includegraphics[width=1\linewidth]{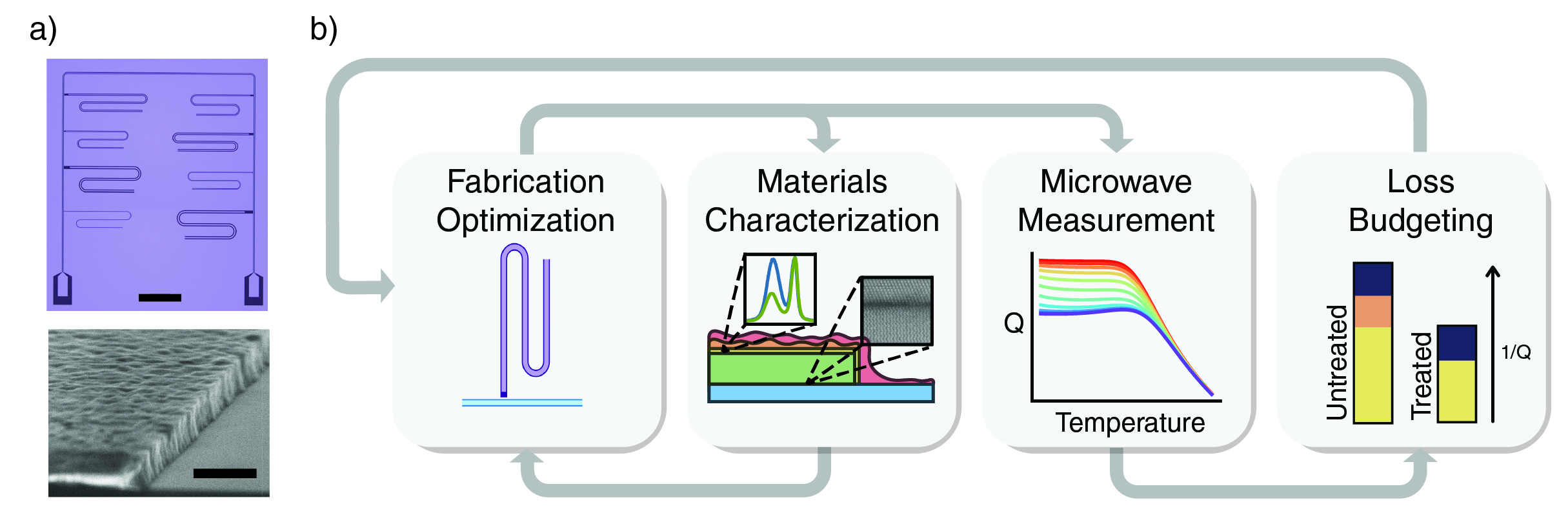}
    \caption{\textbf{Diagnosing sources of loss in superconducting resonators.} a) Top: Optical image of chip containing eight $\lambda$/4 CPW resonators ranging from 2 to 16 $\mu$m in pitch capacitively coupled to a microstrip feedline. Scale bar is 500 $\mu$m. Bottom: Scanning electron microscope image showing sidewall profile of 200 nm Al deposited on c-plane sapphire deposited at 25$\degree$C and fabricated using wet etch. The sidewall is vertical and the film has a lumpy texture. Scale bar is 400 nm. b) Pipeline for measuring materials-related losses using iterative resonator measurement and materials characterization feedback. Microwave measurements provide loss tangents associated with different loss mechanisms, while structural and chemical composition characterization provides candidates for sources of loss. Combined, these measurements allow for estimation of the quantitative contribution of each surface and contaminant to the loss. The cartoon cross section of the superconducting resonator (top) consists of substrate (blue), metal (green), a thin native oxide (yellow), photoresist contamination residue after fabrication (orange), and adventitious hydrocarbons from the atmosphere (red). The metal-substrate (MS) interface and substrate-air (SA) interfaces are represented by a continuous black line. These colors extend to those in the bar graph in the loss budgeting panel.
    }
    \label{fig:fig0}
\end{figure*}

The Josephson junction contains amorphous oxides both on the exposed surfaces of the junction and within the junction tunneling barrier that are known to host two-level systems (TLSs) \cite{zeng2016atomic, fritz2019structural, dubois2013delocalized, lisenfeld2015observation, lisenfeld2019electric}, which give rise to dielectric losses and qubit dephasing. Fabricating Josephson junctions in an ultrahigh vacuum (UHV) ($<$10$^{-9}$ mbar background pressure) environment with a low background pressure of O$_{2}$, H$_{2}$O, and hydrocarbons has already enabled the realization of transmons with lifetime-limited coherence times, demonstrating that the TLS losses associated with Josephson junction fabrication can be mitigated by improved processing and materials optimization \cite{blandbahrami20252D}. In order to study these oxides and evaluate mitigation strategies, we study the power and temperature dependence of losses in aluminum superconducting resonators to quantify the contributions of different sources of loss. We vary fabrication and post-fabrication techniques aimed at decreasing the TLS bath on the surfaces of aluminum coplanar waveguide (CPW) resonators and extract the quality factor limit imposed by linear absorption associated with TLS ($\mathrm{Q_{TLS,0}}$)~\cite{crowley2023disentangling}. We then correlate the chemical, structural, and morphological properties of aluminum resonators with their microwave behavior (Fig.~\ref{fig:fig0}). 

We focus on variations associated with thin film properties, fabrication methods, and post-fabrication processes. We find that aluminum resonators have over three times higher surface dielectric losses than state-of-the-art tantalum resonators. We evaluate the chemical compatibility of various post-fabrication processing steps with aluminum. These losses can be partially offset by device treatment in hydrofluoric (HF) acid, which decreases the losses by a factor of 1.8, approximately matching the performance of tantalum resonators with their native surface oxide and no oxide stripping process \cite{crowley2023disentangling}. However, the growth rate of the native oxide of aluminum is faster than that of tantalum, giving a shorter time window in ambient conditions before the full native oxide thickness has regrown. Based on our studies we estimate that the aluminum oxides associated with the Josephson junction account for $27\%$ of the relaxation rate of state-of-the-art tantalum-on-silicon qubits.

\section{Results}
\begin{figure}
    \centering
    \includegraphics[width=\linewidth]{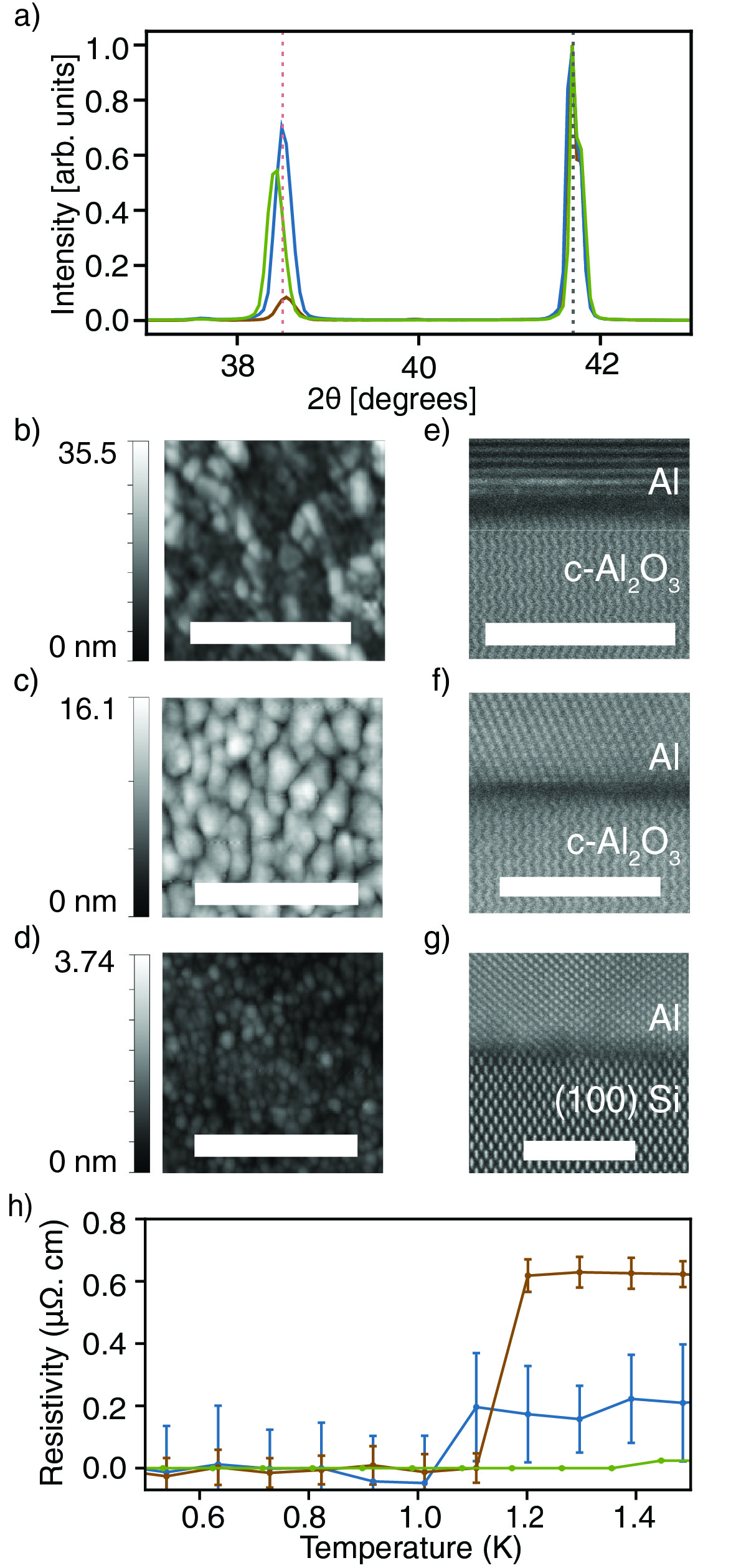}
    \caption{\textbf{Aluminum film characterization.} a) XRD patterns of three aluminum thin films on c-Al$_2$O$_3$ deposited at -72$\degree$C (brown), 25$\degree$C (green) and 200$\degree$C (blue). Dotted lines show Al $<111>$ (2$\theta$ = 38.5$\degree$) and c-Al$_2$O$_3$ $<0006>$ (2$\theta$ = 41.7$\degree$) crystal directions. b) AFM of 200 nm Al film on c-Al$_2$O$_3$ deposited at -72$\degree$C with S$_q$ = 5.11 nm and average grain size of 15.63 nm. c) AFM of 200 nm Al film on c-Al$_2$O$_3$ deposited at 200$\degree$C with S$_q$ = 1.19 nm and average grain size of 87.34 nm. d) AFM of 200 nm Al film on $<100>$ Si deposited at 25$\degree$C with S$_q$ = 0.5 nm and average grain size of 12.58 nm. e) TEM of the interface of Al 200 nm film deposited at  -72$\degree$C with c-Al$_2$O$_3$. Scale bar is 5 nm. f) TEM of the interface of Al 200 nm film deposited at 200$\degree$C with c-Al$_2$O$_3$. Scale bar is 5 nm. g) TEM of the interface of Al 200 nm film deposited at room temperature with $<100>$ Si. Scale bar is 5 nm.  h) Resistivity as a function of temperature for films deposited at -72$\degree$C (brown), 25$\degree$C (green) and 200$\degree$C (blue). }
    \label{fig:fig1}
\end{figure}

We first investigate the structural and dc transport properties of aluminum films grown under different deposition conditions. We tune the deposition rate (0.2 - 0.4~nm/s) and temperature (-72~$\degree$C to 200~$\degree$C), and vary the growth method (e-beam evaporation or sputtering) and substrate (sapphire or Si), (\hyperref[Materials and Methods]{Materials and Methods}). Structural characterization via X-ray diffraction (XRD) and transmission electron microscopy (TEM) diffraction (Online Resource S1 and S2) predominantly show an out-of-plane $<$111$>$ orientation for all deposition conditions. However, the relative intensities of the Al ($<$111$>$) peak at 38.5$\degree$ and the sapphire ($<$0001$>$) peak at 41.7$\degree$ vary as we change the deposition temperature (Fig.~\ref{fig:fig1}a). The Al film deposited at 200~$\degree$C exhibits the highest relative intensity for the $<$111$>$ peak, suggesting a greater degree of crystalline orientation with the substrate. The relative peak intensity is reduced by $\sim$25$\%$ and $\sim$85$\%$ for films deposited at 25~$\degree$C and -72~$\degree$C, respectively. We examine the grain texture of these films with AFM (Fig.~\ref{fig:fig1}b-d), and we observe that the root-mean-square roughness, $S_q$, varies by a factor of $\sim$10 across the three films. Grain sizes are measured with cross-sectional scanning transmission electron microscopy (STEM) and are found to be larger by a factor of $\sim$6.5 for films with a deposition temperature of 200~$\degree$C compared to films grown at -72~$\degree$C and 25~$\degree$C (Table \ref{tab:tab1}, Online Resource S1).

 We also examine the impact of different growth conditions on the heteroepitaxial metal-substrate (MS) interface. STEM images at the MS interface suggest that both Al films grown on sapphire at 200~$\degree$C (Fig.~\ref{fig:fig1}f) and on Si at 25~$\degree$C (Fig.~\ref{fig:fig1}g) feature coherent heteroepitaxy. The Al–substrate interface for films deposited at –72~$\degree$C exhibits blurred contrast observed in the STEM image (Fig.~\ref{fig:fig1}e), indicating reduced heteroepitaxial quality that propagates into the bulk of the Al film. Combining these observations with the observation that the grain sizes for films deposited at -72~$\degree$C and 25~$\degree$C are comparable suggests that the lower intensity peak observed in XRD for the films deposited at -72~$\degree$C arises from a larger out-of-plane $<$111$>$ misalignment (Fig.~\ref{fig:fig1}a and e).  We do not observe any hydrocarbon contamination at the MS interface with energy-dispersive X-ray spectroscopy (EDS) for any film deposition condition, nor do we see evidence of oxygen contamination at grain boundaries across Al films with varying grain sizes (see Online Resource S1).

We next measure the dc resistivity as a function of temperature for each deposition condition to extract the superconducting transition temperature (T$_{c}$) and the residual resistivity ratio (RRR), the ratio of the resistivity at room temperature (T = 300~K) to the resistivity right before T$_{c}$ (T = 2~K) (Fig.~\ref{fig:fig1}h). The highest observed RRR is in films deposited at 200~$\degree$C, RRR = 32$\pm$19, which is larger by a factor of 3 and 6.6 compared to films deposited at -72~$\degree$C and 25~$\degree$C, respectively. This variation in RRR correlates with film morphology, where the highest RRR values are observed for films that have the largest grain sizes. The superconducting transition temperatures are close to the bulk T$_{c}$ of 1.2 K \cite{kittel2004introduction}, ranging from 1.05~K to 1.41~K, and are inversely correlated with grain size.

\begin{table}[H]
\centering
\begin{tabular}{|c|c|c|c|c|}\hline
Film Growth&    Roughness (S$_q$)&  Grain size& T$_c$ (K)&  RRR\\\hline
200$\degree$C&  1.19 nm&    87.34 nm&   1.05 $\pm$ 0.05&   32 $\pm$ 19\\\hline    
25$\degree$C&   0.5 nm& 12.58 nm&   1.41$\pm$ 0.05&   4.82 $\pm$ 0.007\\\hline
-72$\degree$C&  5.11 nm&    15.63 nm& 1.15$\pm$ 0.05& 10.41 $\pm$ 0.7\\ \hline
\end{tabular}
\caption{Aluminum thin film surface roughness, average grain size, T$_c$ and residual resistivity ratio (RRR) for films grown at 200$\degree$C, 25$\degree$C, and -72$\degree$C.}
\label{tab:tab1}
\end{table}

We fabricate 40 microwave resonators across nine chips from Al films with different growth conditions (\hyperref[Materials and Methods]{Materials and Methods}). We use three different fabrication methods: a lift-off method, in which the aluminum is deposited on top of a patterned resist; a wet-etch method, in which the metal is deposited on the substrate and then the patterned resist serves as an etch mask for etching the device in acid; and a dry-etch method, in which the patterned resist serves as an etch mask for reactive ion etching with chlorine chemistry. We note differences in sidewall morphology dependent on etch method (Online Resource S3). All devices are cleaned by iterative solvent processing to remove residual hydrocarbons without damaging the underlying Al (\hyperref[Materials and Methods]{Materials and Methods}).

\begin{figure*}
    \centering
    \includegraphics[width=\linewidth]{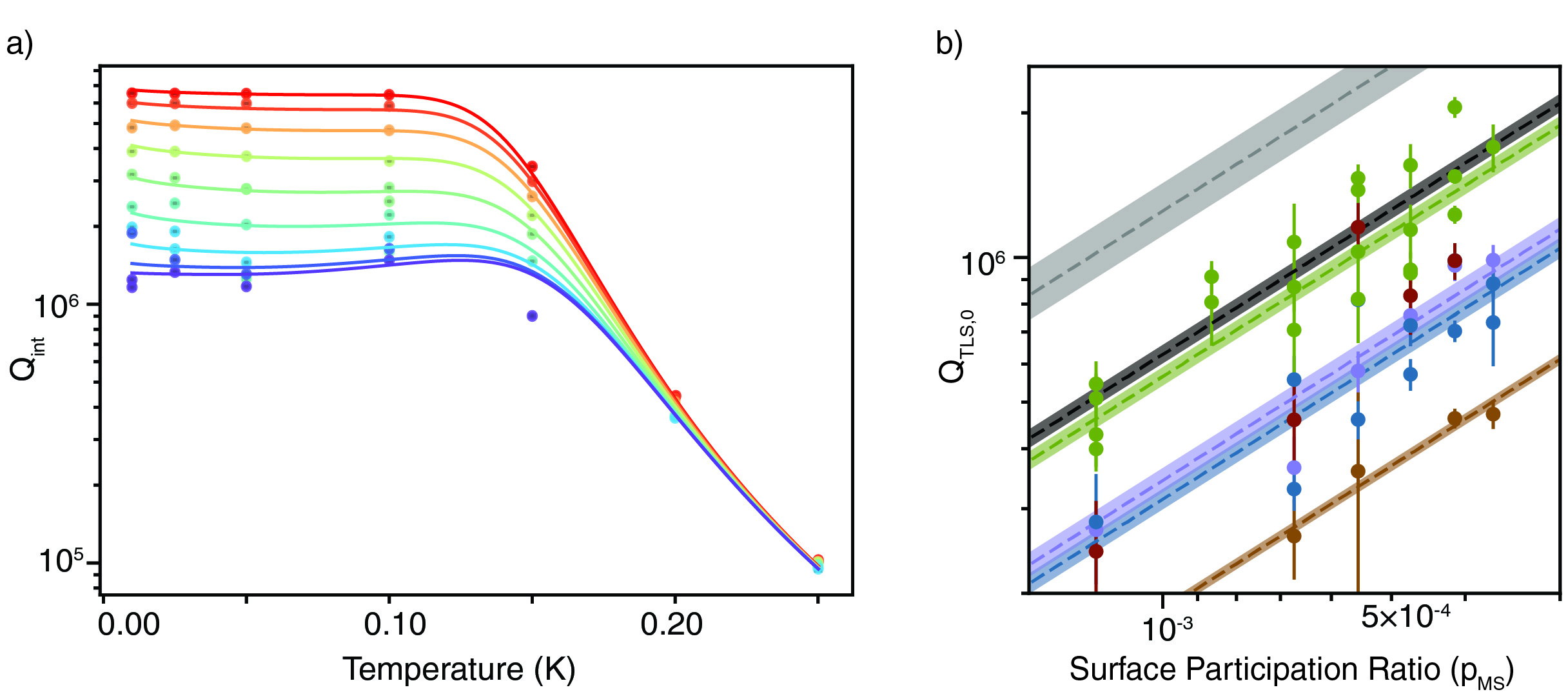}
    \caption{\textbf{Aluminum resonator microwave characterization.} a) Power and temperature dependence of 10 $\mu$m pitch CPW resonator. High power (-10 dB, red) decreases by 10 dB until the lowest power (-90 dB, purple). Three regimes are seen: 1/T dependence until 25 mK, TLS saturation and increase until 125 mK, and quasiparticle rolloff at 150 mK. Scale bar is 500 $\mu$m. b) Q$_\mathrm{{TLS,0}}$ as a function of surface participation ratio. Surface loss tangents are fitted according to surface treatment (tantalum surface loss tangents have been added from Ref. \cite{crowley2023disentangling}); shaded region indicates one standard deviation on fit parameters. From the top, fitted loss tangents are : Ta treated with buffered oxide etch (gray), Ta treated with piranha solution (black), HF-treated Al on sapphire or silicon (green), H$_{2}$O$_{2}$-treated Al on sapphire (purple),  solvent-treated Al on sapphire (blue), and solvent-treated dry etch Al on sapphire, chlorine chemistry (brown). Red points represent HF-treated resonators remeasured after 3 months of oxide and hydrocarbon saturation.}
    \label{fig:fig2}
\end{figure*}

We measure the internal quality factor, Q$\mathrm{_{int}}$, as a function of temperature and microwave power to disentangle the different microwave loss contributions in Al devices (Fig.~\ref{fig:fig2}a) (Online Resource S4). We model Q$_{\mathrm{int}}$ with three separate contributions: Q$_{\mathrm{TLS}}$, the temperature- and power-dependent quality factor due to TLSs; Q$_{\mathrm{QP}}$, the temperature-dependent quality factor due to thermal quasiparticles; and Q$_\mathrm{other}$, a power- and temperature-independent term arising from other loss mechanisms (Fig. \ref{fig:fig2}a)~\cite{crowley2023disentangling}. Similar to tantalum devices in the millikelvin temperature and single-photon power regime~\cite{crowley2023disentangling}, we observe that at least 84$\%$ of total loss in our Al devices is from TLSs. We therefore focus on a fitting parameter that estimates the inverse linear absorption due to TLSs, Q$_\mathrm{{TLS,0}}$. We vary the geometry of the device to examine the dependence of Q$_\mathrm{{TLS,0}}$ on the surface participation ratio (SPR) of the electromagnetic energy density. We find that Q$_\mathrm{{TLS,0}}$ increases as the metal-substrate SPR ($p_{\mathrm{MS}}$) decreases with a linear dependence, indicating that the TLS losses are present in a thin layer at a surface or interface (Fig. \ref{fig:fig2}b, see Online Resource S5). This linear dependence can be interpreted as a dielectric loss tangent (tan $\delta$) associated with a thin surface or interface dielectric layer of uniform thickness, with contributions from the amorphous surface oxide, hydrocarbon contamination, and defects at the metal–air (MA), MS, and substrate–air (SA) interfaces.

We study the variation of this surface loss tangent with differing fabrication methods (Fig. \ref{fig:fig2}b, Table \ref{tab:tab2}). The surface loss tangents associated with wet-etch and lift-off fabrication methods are statistically indistinguishable (within 2$\sigma$), and combining the data from both fabrication methods yields a surface loss tangent tan $\delta$ = (3.19 $\pm$ 0.22) $\times$ 10$^{-3}$, while the dry-etch method yields a surface loss tangent of tan $\delta$ = (5.43 $\pm$ 0.14) $\times$ 10$^{-3}$ (Fig.~\ref{fig:fig2}b, brown data), corresponding to a $\sim$70\% increase in TLS loss. Overall our measurements suggest that resonators fabricated from films deposited at 200~$\degree$C with the wet etch fabrication method result in the best observed device performance; therefore, we use this fabrication procedure for studying the impact of post-fabrication surface treatments.

\begin{table*}
    \centering
    \begin{tabular}{|c|c|c|c|c|c|}\hline
         Chip&   Film Deposition&Fabrication method& Surface treatment &Loss Tangent ($\times 10^{-3}$)&Oxide Thickness \\\hline
         Wet etch&  200$\degree$C, c-Al$_2$O$_3$&   Wet Etch&   untreated& 2.53 $\pm$ 0.14 &   2.69 $\pm$ 0.07 nm\\\hline
         H$_2$O$_2$&    200$\degree$C,  c-Al$_2$O$_3$&  Wet Etch&    H$_2$O$_2$& 2.91 $\pm$ 0.17 &   2.66 $\pm$ 0.07 nm\\\hline
         Cryo&  -72$\degree$C,  c-Al$_2$O$_3$&  Liftoff&    untreated&   3.60 $\pm$ 0.21 &   2.73 $\pm$ 0.08 nm\\\hline
         Dry etch&   200$\degree$C, c-Al$_2$O$_3$&  Reactive Ion Etch&  untreated&    5.43 $\pm$ 0.14 &   2.70 (est.) nm\\\hline
         H$_2$O$_2$, 49\% HF 5s& 200$\degree$C, c-Al$_2$O$_3$& Wet Etch&    49\% HF&    1.86 $\pm$ 0.18&    1.90 $\pm$ 0.05 nm\\\hline
          49\% HF, 3 months&  200$\degree$C, c-Al$_2$O$_3$&   Wet Etch&   HF, oxide saturation&  2.51 $\pm$ 0.29 &   3.11 $\pm$ 0.09 nm\\\hline
         49\% HF 15s&   200$\degree$C, c-Al$_2$O$_3$&    Wet Etch& 49\% HF& 1.68 $\pm$ 0.14 & 1.91 $\pm$ 0.05 nm\\\hline
         RT, Si, 49\% HF 10s&    25$\degree$C, 100 Si&   Wet Etch&   49\% HF&  1.34 $\pm$ 0.07&    1.87 $\pm$ 0.05 nm\\ \hline
 RT, Si, 49\% HF 10s& 25$\degree$C, 100 Si& Wet Etch& 49\% HF& 1.95 $\pm$ 0.08& 2.00 (est.) nm\\\hline
    \end{tabular}
    \caption{Individual resonator chip fabrication methods, growth parameters, surface treatments, loss tangents and oxide thicknesses for all devices. Devices listed with estimated (est.) oxide thicknesses are found using the time exposed to air before cryogenic measurement and the corresponding oxide thickness in Fig. \ref{fig:fig3}a. }
    \label{tab:tab2}
\end{table*}

\begin{figure*}
    \centering
    \includegraphics[width=\linewidth]{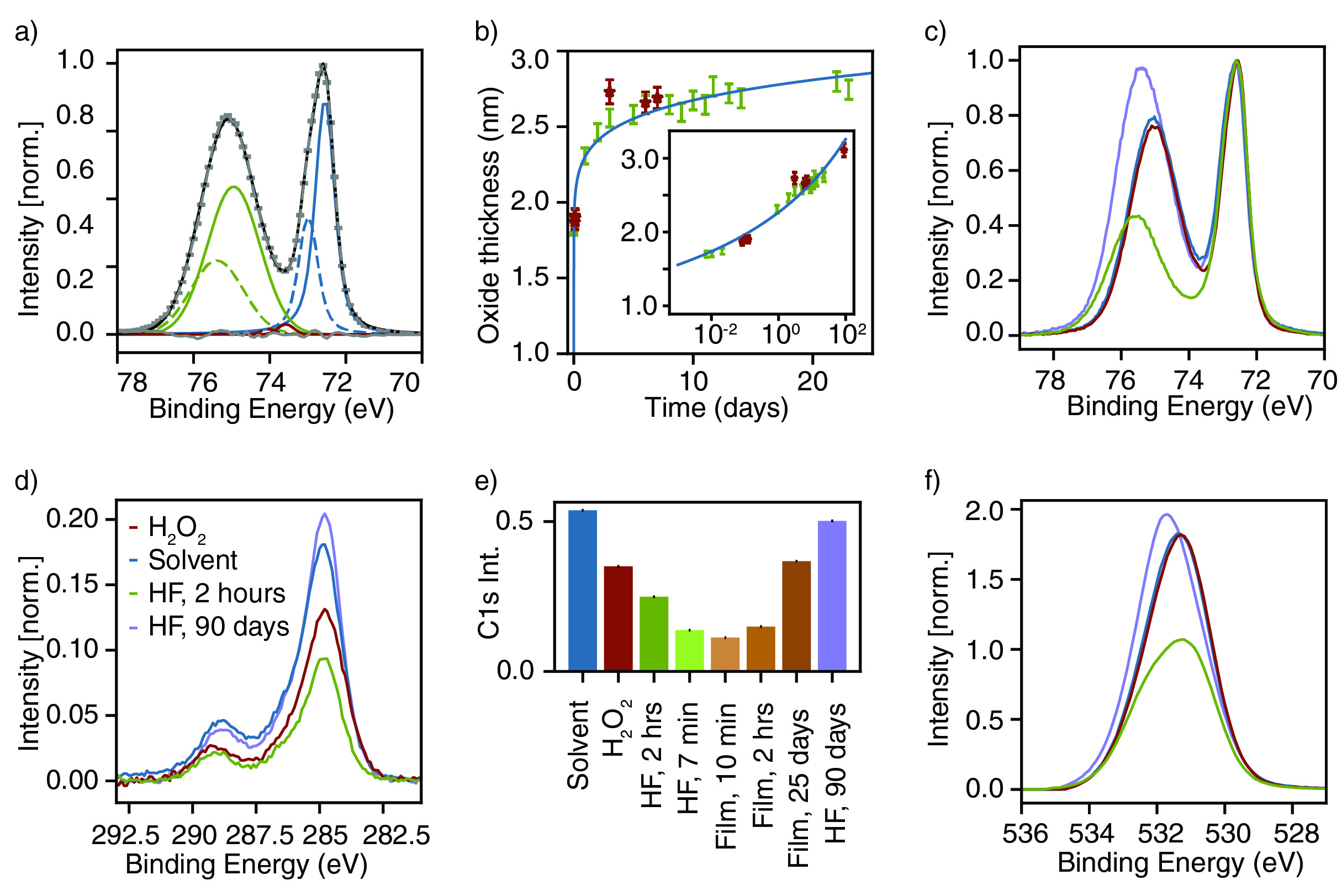}[H]
    \caption{\textbf{Aluminum surface characterization using X-ray Photoelectron Spectroscopy.} a) Example fit of Al2p peak for a film with 2.69 $\pm$ 0.07nm of native oxide. Each oxidation state (0, 3+, interface) has a 1/2 (dotted) and 3/2 (solid) spin orbital contribution. Using the intensity of oxide peaks to the metal peak, we calculate the oxide thickness using the Strohmeier equation (Online Resource S6). Plot is normalized to intensity. b) Oxide thicknesses calculated for a native oxide on aluminum film (green) as a function of time. The fitted line (blue) shows a linear regime with rapid oxide growth kinetics for roughly the first 12 hours, then an intermediate regime until 24 hours after exposure to air with slower, but still relatively fast growth, followed by a logarithmic dependence which slowly saturates the oxide. Points in red show oxide thicknesses measured from devices. Inset: The same data plotted on semilog axes.c) Al2p XPS intensity as a function of binding energy. H$_2$O$_2$ and solvent-cleaned devices show 2.66 $\pm$ 0.07 and 2.69 $\pm$ 0.07 nm of oxide growth, respectively, while the HF-cleaned device shows 1.87 $\pm$ 0.05 nm of growth.  Plots are normalized to intensity, and then scaled so the Al2p peak height is set to 1 to more easily see differences in oxide thickness. All spectra are shifted to the Al$^0_{3/2}$ = 72.6 eV. d) C1s XPS intensity as a function of binding energy normalized to the corresponding intensity of the Al2p peak. The solvent-cleaned device shows a greater intensity than H$_2$O$_2$, which shows a greater intensity than the HF-cleaned device. e) Numerical comparisons of the integral intensities of the C1s peak across devices and films at each of the growth regimes. f) O1s XPS intensity as a function of binding energy normalized to the corresponding intensity of the Al2p peak. The HF-treatment produces a double peaked structure with a small increase in binding energy, signifying a slight decrease in interfacial oxide between the metallic and 3+ aluminum states. We normalize the peak positions for both Al2p and O1s to the C1s peak at 284.8~eV and peak intensities for both C1s (Fig. \ref{fig:fig3}d) and O1s (Fig. \ref{fig:fig3}f) to the corresponding Al2p intensity to accurately compare carbon contamination between different post-fabrication cleaning treatments. }
    \label{fig:fig3}
\end{figure*}

We use two different post-fabrication surface treatments to mitigate surface losses: a 49\% HF acid treatment, and a hydrogen peroxide (H$_{2}$O$_{2}$) treatment. We evaluate the two post-fabrication surface treatments on the amorphous aluminum oxide and fabrication hydrocarbon residue by comparing the variation in Al2p, O1s, and C1s peaks in X-ray photoelectron spectroscopy (XPS) (Online Resource S6). The Al2p spectrum consists of a pair of peaks split by spin-orbit coupling associated with metallic Al at $\sim$72.6~eV and another pair of peaks associated with aluminum oxide at $\sim$75~eV. We model the Al2p peaks with two Lorentzian lineshapes for the metallic peaks (Al$^{0}$) and Gaussian lineshapes for the non-zero oxidation states (Al$\mathrm{_{int}}$, Al$^{3+}$) (Fig. \ref{fig:fig3}a). We estimate the oxide thickness from the integrated oxide peak intensity as a function of time in atmosphere over 25 days, and we observe oxidation kinetics characterized by two distinct regimes: a linear regime from 0 - 2.3~nm over the course of 24 hours, followed by a logarithmic dependence thereafter which saturates at around 3 nm, consistent with the Cabrera-Mott model (Fig. \ref{fig:fig3}b)~\cite{cabrera1949theory, ramirez2022testing, cai2011tuning, baran2014mechanism, lousada2018first, lanthony2012early}.

We use 49\% HF acid treatment to reduce the amorphous aluminum oxide layer and eliminate hydrocarbon contamination arising from photoresist residue. Comparing a film that has been treated with 49\% HF to a reference sample that has been freshly deposited with the same duration of ambient exposure, the aluminum oxide thickness and surface carbon contamination are similar as measured by XPS, indicating that this surface treatment completely removes aluminum surface oxide until re-exposure to air (Fig. \ref{fig:fig3}c, d and e). We minimize the device exposure to atmosphere before cooldown; the device is exposed to ambient conditions for 2-4 hours resulting in a 1.7-2.1~nm aluminum oxide layer thickness (Fig. \ref{fig:fig3}a). The surface loss tangent for the devices with 49\% HF acid treatment shows a statistically significant ($>$3$\sigma$) improvement compared to untreated devices; tan $\delta$ = (1.77 $\pm$ 0.08) $\times$ 10$^{-3}$ (Fig.~\ref{fig:fig2}b, green data), corresponding to a $\sim$40\% decrease in surface loss (Table~\ref{tab:tab2}). 

Hydrogen peroxide (H$_{2}$O$_{2}$) treatment decreases the intensity of the surface hydrocarbon content as measured by the C1s spectrum, while the aluminum oxide thickness and binding energy of the O1s peak remain unaffected (Fig. \ref{fig:fig3}d, e, f). This post-fabrication treatment reduces the surface hydrocarbons including fabrication and adventitious hydrocarbon residues by (34.7 $\pm$ 5.8)$\%$; however, we observe no statistically significant improvement, ($<$1$\sigma$), compared to the surface loss tangent associated with the wet-etch and lift-off fabrication methods with no post-fabrication treatment. Although we observe $\sim$34.7$\%$ reduction in the carbon signal on the top surface of the Al device in XPS, we cannot directly assess the differential effectiveness of this treatment on different types of hydrocarbon contamination, namely fabrication residue and adventitious carbon from the atmosphere, using XPS alone. Since the surface losses are unchanged, we conclude that H$_{2}$O$_{2}$ treatment removes adventitious carbon, but not fabrication residue, and that the latter has a larger contribution to surface loss.

\section{Discussion}

The comparison between devices without post-fabrication processing and with 49\% HF acid treatment suggests that the dominant source of TLS loss in Al devices is the amorphous aluminum oxide at the MA interface. The surface loss tangent for devices without any post-fabrication treatment contains loss contributions from the MA (which includes amorphous aluminum oxide, $\mathrm{\tan \delta _{AlO_x}}$) and hydrocarbon contamination ($\tan \delta_{\mathrm{HC}}$)), SA ($\tan \delta_{\mathrm{SA}}$), and MS ($\tan \delta_{\mathrm{MS}}$) interfaces. On the other hand, for the 49\% HF acid-treated devices we significantly reduce the amorphous surface oxide and hydrocarbon contamination by $\sim$29.6$\%$ and $\sim$53.6$\%$, respectively, and expect the remaining hydrocarbons to be adventitious rather than fabrication residue. We therefore model the total surface loss tangent for these devices with respect to each interface SPR ($p_{\mathrm{i}}$) and estimated dielectric layer thickness (t$_{i}$) normalized to the assumed thickness (t$_{0}$ = 3nm) in our SPR simulation as follows (Online Resource S7):

\begin{align}
\label{equation1}
    \tan \delta_\mathrm{Al, HF} = &\mathrm{\frac{p_{MA}}{p_{MS}}}  \cdot  \mathrm{\frac{t_{HF, AlO_x}}{t_0}} \cdot \tan \delta_\mathrm{{AlO_x}} \\ 
    & \nonumber + \mathrm{\frac{p_{SA}}{p_{MS}}} \cdot \mathrm{\frac{t_{SA}}{t_0}} \cdot  \tan \mathrm{\delta_{SA} + \frac{t_{MS}}{t_0}} \tan \delta_\mathrm{MS}\\
    \label{equation2}
    \tan \delta_\mathrm{Al, HF 90 days} = & \mathrm{\frac{p_{MA}}{p_{MS}}}  \cdot  \mathrm{\frac{t_{HF 90 days, AlO_x}}{t_0}} \cdot \tan \delta_\mathrm{{AlO_x}} \\
    & \nonumber + \mathrm{\frac{p_{SA}}{p_{MS}}} \cdot \mathrm{\frac{t_{SA}}{t_0}} \cdot  \tan \mathrm{\delta_{SA} + \frac{t_{MS}}{t_0}} \tan \delta_\mathrm{MS}\\
    \label{equation3}
    \tan \delta_\mathrm{Al, untreated} = & \mathrm{\frac{p_{MA}}{p_{MS}}}  \cdot  \mathrm{\frac{t_{untreated, AlO_x}}{t_0}} \cdot \tan \delta_\mathrm{{AlO_x}} \\ 
    & \nonumber +  \mathrm{\frac{p_{MA} + p_{SA}}{p_{MS}} } \cdot \mathrm{\frac{t_{i, HC}}{t_0}} \tan \delta_\mathrm{HC} \\  
    \nonumber
    & + \mathrm{\frac{p_{SA}}{p_{MS}}} \cdot  \mathrm{\frac{t_{SA}}{t_0}} \cdot \tan \mathrm{\delta_{SA}
    + \frac{t_{MS}}{t_0}} \tan \delta_\mathrm{MS}
\end{align}

We isolate the loss contribution from amorphous aluminum oxide, $\tan \delta_{\mathrm{AlO_{x}}}$, by comparing the extracted surface loss tangents and the surface oxide thickness for the 49\% HF treated aluminum after atmospheric exposure for 2 hours to the same film treatment after exposure for 90 days (Fig.~\ref{fig:fig3}c and d, purple data points). We confirm that the oxide thickness and adventitious hydrocarbon signal have saturated after this 90 day period via XPS analysis (see Online Resource S7). We are confident that the MS and SA interfaces of these devices are the same, as they are deposited and fabricated using the same process; therefore, the difference between their loss tangents can be entirely attributed to the change in oxide thicknesses, equal to 1.2 nm. 

We therefore estimate the intrinsic material loss tangent of amorphous AlO$_{x}$ to be $\tan \delta_{\mathrm{AlO_{x}}} = (1.74 \pm 0.7) \times 10^{-2}$, which can be scaled by surface participation ratios $(p_{\mathrm{MA}}/{p_{\mathrm{MS}}})\tan \delta_{\mathrm{AlO_{x}}} = (1.82 \pm 0.74) \times 10^{-3}$. This value is over three times larger than what is estimated for TaO$_{x}$, $\tan \delta_{\mathrm{AlO_{x}}} = (5 \pm 1) \times 10^{-3}$ ($(p_{\mathrm{MA}}/{p_{\mathrm{MS}}})\tan \delta_{\mathrm{TaO_{x}}} = (6 \pm 1) \times 10^{-4}$) ~\cite{crowley2023disentangling}. 

Using Eq. \ref{equation1}, we can insert $\mathrm{\tan \delta_ {AlO_x}}$ to find that the combined, surface-scaled contribution from the SA and MS interfaces is $(6.19 \pm 4.96) \times 10^{-4}$.  This is $\sim$ 50\% higher than the TLS loss from the same interfaces in tantalum on sapphire devices \cite{crowley2023disentangling}.

We evaluate the loss contribution from the SA interface to the total surface loss by comparing the resonators measured on sapphire and Si substrates. Sapphire is a chemically inert single crystal, whereas Si can regrow an amorphous surface oxide layer after exposure to air. The 49\% HF acid-treated devices are fabricated on both sapphire and Si substrates (Fig.~\ref{fig:fig2}b, green data points). The 49\% HF acid treatment removes the amorphous silicon oxide layer, leaving a $\sim$0.4 nm thick SiO$_{x}$ layer on the SA interface of the Si devices (Online Resource S7). The surface loss tangent for Al-on-Si devices, however, is within 1$\sigma$ of their sapphire counterpart devices, suggesting SiO$_{x}$ has less than 5\% contribution to the total surface loss, similar to what has been reported for Ta devices \cite{blandbahrami20252D}. 

We next examine the MS interface contribution to the total surface loss by comparing the devices fabricated with the liftoff method (film deposited at -72~$\degree$C) to the wet-etch method (film deposited at 200~$\degree$C). While the heteroepitaxial interfaces differ greatly in imaging and structural analysis, the extracted surface loss tangents display comparable behavior to one another (within 1$\sigma$). Similarly, devices with the 49\% HF acid treatment on sapphire (film deposited at 200~$\degree$C) and Si (film deposited at 25~$\degree$C) also exhibit similar losses. This observation confirms the variations in crystalline orientation, morphology, and heteroepitaxy quality at the MS interface have a negligible contribution to the surface loss for Al resonators.

We observe a peak in the C1s spectrum at 284.8~eV for all devices, with the largest intensity corresponding to untreated devices (Fig.~\ref{fig:fig3}d, untreated). This signal includes fabrication residue from photoresist, as well as adventitious carbon from atmospheric exposure. The 49\% HF acid treatment strips away the surface oxide and therefore completely eliminates the fabrication contamination residue. We interpret the remaining C1s signal as related to atmospheric adventitious carbon (Fig.~\ref{fig:fig3}d, HF, 2 hours). Using the estimated loss tangent for AlO$_{x}$ in Eq.~\ref{equation1}, and assuming that the photoresist residue resides on the MA and SA surfaces, we estimate a loss tangent from photoresist residue to be $\mathrm{\tan\delta_{HC} = (3.9 \pm 1.1) \times 10^{-3}}$ ($\mathrm{(p_{MA}+p_{SA})/p_{MS} \times \tan\delta_{HC} =  (4.9 \pm 1.4) \times 10^{-3}})$ . 
We see the difference between adventitious carbon and fabrication residue starkly when we compare the untreated aluminum device loss tangent with the 49 \% HF treated aluminum loss tangent that was left to saturate in both hydrocarbons and oxide over 90 days. The 49\% HF acid-treated device after exposure to atmosphere for 90 days shows a C1s intensity $\sim$8\% less than the signal observed for untreated devices (Fig.~\ref{fig:fig3}e). However, the sources of hydrocarbon contamination are different in each case: the untreated devices are mainly contaminated by fabrication hydrocarbon residue, and the 49\% HF acid-treated device after exposure to atmosphere for 90 days contains only adventitious hydrocarbons.  Despite a similar hydrocarbon coverage estimated by XPS, the surface loss tangent for the device treated with 49\% HF and left in atmosphere for 90 days has 21\% less TLS loss despite having 15\% more oxide. Comparing these surface loss tangents suggests that the loss contribution from the adventitious hydrocarbons must be much lower than that of fabrication residue, and is therefore negligible compared to the TLS losses for aluminum oxide, fabrication residue, and the SA and MS interfaces.

We combine the measured losses at each interface to estimate a loss budget for aluminum resonators. For a co-planar waveguide resonator with a 10 $\micro$m gap fabricated from aluminum with no further surface processing, we estimate aluminum oxide to account for $ (52.4 \pm 21.5)\%$, hydrocarbons for $(27.7 \pm 7.2) \%$, and the MS and SA surfaces combined for $(19.8 \pm 15.9) \%$ of surface TLS losses. Furthermore, we can estimate the contribution of surface TLS loss in the aluminum Josephson junction leads to the total loss in state-of-the-art Ta-on-Si transmon qubits \cite{blandbahrami20252D}. We simulate the electric field energy density of the resonant qubit mode while ignoring a small radius around the junction barrier oxide because of the difficulty of obtaining accurate electric field simulations at the tunnel barrier. We then use the measured loss tangents at a microwave power that corresponds to one intracavity photon ($ \mathrm{\bar n = 1}$), the relevant parameter range for qubit operation, to compute the contribution of each interface to the total qubit relaxation rate. We note that at $ \mathrm{\bar n = 1}$, the surface loss tangent for $\mathrm{AlO_x}$, $\mathrm{\tan \delta_{AlO_x, \bar n = 1}}$, is reduced by a factor of five compared to linear absorption ($ \mathrm{\bar n = 0}$) (Online Resource S7).  We find that 19.7\% of the TLS loss in state-of-the-art transmon qubits with the geometry reported in \cite{blandbahrami20252D} arises from the surfaces of the Al leads associated with the Josephson junction, while 73.5\% is attributed to the Ta-based capacitor pads. We then compare the simulated qubit shunt capacitance with the calculated parallel-plate capacitance of the junction and solve for the remaining TLS loss arising from the junction barrier itself, and attribute 6.8\% of the total TLS loss to the barrier of the Josephson junction. Notably, the aluminum leads have an effective Q of 48 million, thereby representing an important target for loss mitigation.

Comparing these loss estimates to the performance of Ta-on-Si qubits, we estimate that the tunnel barrier itself must have a Q greater than 142 million, which implies that the loss tangent of the aluminum oxide junction barrier, $\tan\delta_\mathrm{AlO_{x} barrier, \ \bar n=1} = (4.3 \pm 3.7) \times 10^{-7}$, is four orders of magnitude lower than the surface oxide $\mathrm{\tan \delta_{AlO_x, \bar n = 1}}$. This low tunnel barrier loss tangent estimate is consistent with previous estimates of the tunnel barrier loss tangent in ``mergemon" geometries \cite{mamin2021merged}. A likely source for this discrepancy is a low spectral and areal density of TLSs leading to the complete absence of strongly-coupled TLSs in the junction. In the geometry used in Ref. \cite{blandbahrami20252D}, the junction has dimensions $\mathrm{A_{JJ, dielectric} = 200 \ nm \times 200 \ nm = 4 \times 10^{-2} \micro m^2}$, 
$\mathrm{V_{JJ, dielectric} = A_{JJ, dielectric} \times 2 \ nm = 8 \times 10^{-5} \micro m^3}$. In junctions with comparable area, values of $\sim$ 0.5-1.5 TLS (GHz $\micro m^2)^{-1}$  and 250-760 TLS (GHz $\micro m^3)^{-1}$ have previously been reported \cite{osman2023mitigation, martinis2005decoherence, kline2009josephson, lisenfeld2019electric, bilmes2022probing}. This corresponds to a predicted 0.02 to 0.06 TLS per GHz for a 200 nm by 200 nm area (200 nm by 200 nm by 2 nm volume) junction. The low observed loss tangent also implies that the primary source of microwave loss is TLSs.  Therefore avoiding strongly coupled TLS is sufficient to achieve a low-loss junction tunnel barrier in individual qubits, though improvements in TLS density will be needed to ensure consistent performance across all qubits in large-scale processors.

We lastly compare the TLS loss in Al resonators to Ta resonators. Ta is chemically robust, a property that allows for more aggressive post-fabrication processes such as piranha solution (2:1 H$_{2}$SO$_{4}$: H$_{2}$O$_{2}$) and buffered oxide etch (BOE), which reduces surface oxides with slow oxide regrowth kinetics and eliminates the fabrication hydrocarbon contamination.  These properties enable state-of-the-art tantalum resonators with loss tangents of (1.59 $\pm$ 0.07) $\times\mathrm 10^{-3}$ (Fig.~\ref{fig:fig2}b, black data) and  (8.1 $\pm$ 0.6) $\times\mathrm 10^{-4}$ (Fig.~\ref{fig:fig2}b, gray data), respectively~\cite{crowley2023disentangling}. The 49\% HF acid-treated Al resonators perform within 1$\sigma$ of the piranha-treated Ta devices; however, BOE-treated Ta devices show two times lower TLS loss than HF-treated Al devices. To achieve comparable surface losses to BOE-treated Ta, the aluminum oxide thickness would need to be decreased to about 0.5 nm, which is incompatible with procedures for standard packaging and wiring in ambient conditions because of aluminum's fast oxide growth kinetics.

\section{Conclusion}

In this study, we use microwave measurements of Al resonators and correlate them with detailed materials characterization to quantify loss mechanisms. We find that the dominant source of loss is TLSs associated with the surface aluminum oxide, which is four times more lossy than the surface oxide of Ta in state-of-the-art devices. This loss can be mitigated by treatment with 49\% HF, which reduces the oxide thickness by $\sim$0.8~nm and thereby reduces the surface losses by a factor of 1.8; however, rapid surface oxide regrowth makes this surface condition difficult to translate to large-scale processors. Our results also suggest that differences in film crystallinity and morphology have a small contribution to the total surface loss compared to aluminum oxide and hydrocarbon contamination related to fabrication residue. Our measurements of aluminum and aluminum oxide allow us to estimate that they contribute around $27\%$ of the relaxation rate of state-of-the-art Ta qubits  \cite{blandbahrami20252D} (Online Resources S7 and S8). These results underscore the importance of surface cleaning and surface oxide mitigation in continued improvement of transmon qubits. Ongoing work targets the suppression of surface oxide regrowth in Al devices via full encapsulation and the optimization of post-fabrication cleaning to mitigate fabrication residues and surface oxides near the Josephson junctions of Ta qubits. Furthermore, the TLS loss associated with the surface of Al is a potential source of temporal fluctuations and dephasing from strongly-coupled TLSs \cite{zanuz2025mitigating, lisenfeld2019electric, bilmes2022probing}, and eliminating these TLSs will be important for improving the performance of large-scale superconducting quantum processors.

\section{Materials and Methods}
\label{Materials and Methods}
\subsection{Film growth}
\label{section:A}
Prior to film deposition of heated films, we clean 1" by 1" pieces of c-plane HEMEX grade sapphire substrates (Crystal Systems) in a 2:1 H$_2$SO$_4$:H$_2$O$_2$ piranha solution to rid the substrate surface of hydrocarbon residue. We rinse the substrates in DI water three times, submerged in IPA and blow dried with a nitrogen gun. We load the sapphire into the loadlock of an e-beam evaporator (Plassys MEB550S) and transfer to the deposition chamber after reaching a vacuum of $\mathrm{5 \times 10^{-7}}$ mTorr in the loadlock. We heat the sapphire to 300$\degree$C in situ for one hour, then cool to 200$\degree$C for deposition. We deposit aluminum films from an unlined copper crucible at a rate of 0.25 nm/s and a thickness of 200$\pm$10 nm as measured by a quartz crystal monitor at a base pressure of $\mathrm{2.9 \times 10^{-10}}$ mTorr. After deposition, we let films cool to ambient temperature in vacuum over the course of 12 hours.

For cryogenically deposited films, the substrate with patterned resist is loaded into the loadlock, transferred to the deposition stage and cooled using a liquid N$_2$ cold finger which directly contacts the substrate holder. When a steady state temperature is reached (-72$\degree$C), the cold finger is retracted and aluminum is deposited using the same specifications as above. Films are left to warm in vacuum to ambient temperature over the course of 12 hours.

For the aluminum film deposited at room temperature on high-resistivity silicon ($>$ 20 k$\ohm$ cm, Siegert), we use piranha solution (same as above) to remove hydrocarbons and oxidize the surface. Next, we submerge the silicon substrate in a 10:1 DI water:HF solution for 3 minutes to remove the surface oxide. After rinsing 3 times with DI water and drying completely with nitrogen, we transfer silicon to the Plassys loadlock within 10 minutes to ensure no interfacial oxide \cite{blandbahrami20252D}. We evaporate the Al film at a rate of 0.4 nm/s and transfer the substrate out immediately after finishing.
\subsection{Device fabrication}
\label{section:D}
For devices with subtractive etching (wet etching or reactive ion etching), we grow aluminum films prior to fabrication. We prepare the surface with hexamethyldisilazane (HMDS) at 148$\degree$C in a YES vapor prime oven to promote adhesion of the resist onto the surface. We spin AZ1518 onto the film at 4000 rpm for 40 seconds then bake at 95$\degree$C for one minute on a hotplate. We then use direct-write lithography (Heidelberg DWL66+) to define the resonator pattern using the 10mm writehead. We hard bake the resist on a hot plate at 110$\degree$C for two minutes, then develop using AZ300MIF developer for 90 seconds, followed by a 30s rinse in deionized (DI) water and drying using a nitrogen gun.

We use both wet etching and reactive ion etching (RIE) to define device features. For wet etched devices, we place samples in a mixture of 55-65\% phosphoric acid, 1-5\% nitric acid, 5-10\% acetic acid and balanced DI water (Transene Aluminum Etchant Type A) for 7-10 minutes at room temperature, removing the film roughly one minute after the features become visible. We rinse the device in DI water three times and blow dry with a nitrogen gun. For dry etch devices, we use a chlorine-based chemistry in an inductively coupled plasma reactive ion etcher (PlasmaTherm Takachi SLM) with a pressure of 4 mTorr, flow rate of 1.5 sccm Cl$_2$, 3 sccm BCl$_3$, and 25 sccm Ar, bias power of 70W, RF power of 300W at 25$\degree$C for 75s. 

For samples fabricated using liftoff, we employ the same HMDS prime directly on the sapphire substrate. We spin a bilayer resist stack of LOR 3A and AZ1518 to achieve a small undercut to ensure clean liftoff. We spin LOR 3A at 4000 rpm for 40s and bake for 5 minutes at 160$\degree$C, followed by AZ1518, spun and baked with the above parameters. Photolithography follows the same procedure as above, but the AZ300MIF development is 2 minutes long to achieve an undercut in the bilayer. After rinsing in DI water and drying using a N$_2$ gun, we finish with a five minute, 500W O$_2$ ashing step (Tepla IoN O$_2$ Etcher) to clean the substrate and improve adhesion of the metal layer.  

For all samples, we remove resist using Remover PG at 80$\degree$C for two hours, followed by sonication in acetone and IPA for 5 minutes and drying using a nitrogen gun. We then clean devices using a solvent series, 30\% H$_2$O$_2$, or 49\% HF, and subsequently wirebond and package (QDevil QCage.24).

\subsection{Surface cleaning}
\label{section:F}
Three different cleanings were used to study surface losses on aluminum devices: a solvent series, 30\% H$_2$O$_2$ or 49\% HF. The solvent series was designed to minimize hydrocarbon residue while not damaging the underlying aluminum and consisted of two cycles of various polar aprotic, polar protic and nonpolar solvents, where each solvent is miscible with the previous one. The exact cycle is as follows, where the device is sonicated in each solvent for 20 minutes unless specified: DI water, acetone, pentane, methanol, acetone, toluene, acetone (5 min), IPA (5 min). The chip is finally nitrogen dried and packaged immediately.

30\% H$_2$O$_2$ was utilized to more aggressively target hydrocarbons after the solvent series showed noticeable carbon contamination in XPS. Devices were placed into roughly 30 mL of 30\% H$_2$O$_2$ and left for 30 minutes. After H$_2$O$_2$, the device was rinsed three times in DI water, sonicated in IPA for 5 minutes and blow dried with nitrogen.

49\% HF was utilized to fully remove the surface oxide and any associated surface contaminants. The device was submerged in 49\% HF for 5-20 seconds and then quickly rinsed into successive DI water 5 times, then sonicated for 5 minutes in IPA and blow dried with nitrogen. After HF treatment on sapphire substrates, we observed variability in the uniformity of aluminum oxide etching, which we attributed to non-uniformity of the starting aluminum film surface, such as an unsaturated oxide or micro-masking from hydrocarbon residue that was still present on the device.  Because this treatment is only 5-10s, small changes in the surface chemistry such as these can lead to visible variations in the aluminum surface after the treatment. After seeing this issue, we took steps to saturate the oxide and remove hydrocarbons prior to the HF treatment by implementing a five minute, 500W O$_2$ ashing step (Tepla IoN O$_2$ Etcher) before submerging Al on Si devices in HF, as we iterated on sapphire to first test the feasibility of the HF treatment before attempting to do so on silicon. We verified in XPS that all residual fabrication hydrocarbons were removed and that the oxide was slightly grown.

\section{Acknowledgments}
We acknowledge Lev Krayzman, Sounak Mukherjee, ad Bert Harrop for helpful discussions.  
This work was primarily supported by the U.S. Department of Energy, Office of Science, National Quantum Information Science Research Centers, Co-design Center for Quantum Advantage (C$^{2}$QA) under Contract No. DESC0012704. This material is based upon work supported by Google Quantum AI under SOW No. 89201. This material is based upon work supported by the Air Force Office of Scientific Research under award number FA9550-25-1-0172. E. H. and M. P. B. were also supported by the National Science Foundation Graduate Research Fellowship Program (NSF GRFP) under grant No. DGE-2444107. The authors acknowledge the use of Princeton’s Imaging and Analysis Center (IAC), which is partially supported by the Princeton Center for Complex Materials (PCCM), a National Science Foundation (NSF) Materials Research Science and Engineering Center (MRSEC; DMR-2011750), as well as the Princeton Micro/Nano Fabrication Laboratory.

Princeton University Professor Nathalie de Leon is a Visiting Faculty Researcher with Google Quantum AI. Due to her income from Google, Princeton University has a management plan in place to mitigate a potential conflict of interest that could affect the design, conduct, and reporting of this research. Her academic group also has a sponsored research contract with Google Quantum AI. Princeton University Professor Andrew Houck is also a consultant for Quantum Circuits Incorporated (QCI). Due to his income from QCI, Princeton University has a management plan in place to mitigate a potential conflict of interest that could affect the design, conduct, and reporting of this research.

\newpage

\bibliographystyle{unsrtnat}
\bibliography{biblio}

\end{document}


\title{Supplementary Information for: ``Quantifying surface losses in superconducting aluminum microwave resonators"}

\maketitle

\author{Elizabeth Hedrick}\affiliation{\ECEPrinceton}
\author{Faranak Bahrami}\affiliation{\ECEPrinceton}
\author{Alexander C. Pakpour-Tabrizi}\affiliation{\ECEPrinceton}
\author{Atharv Joshi}\affiliation{\ECEPrinceton}
\author{Q. Rumman Rahman}\affiliation{\ECEPrinceton}
\author{Ambrose Yang}\affiliation{\ECEPrinceton}
\author{Ray D. Chang}\affiliation{\ECEPrinceton}
\author{Matthew P. Bland}\affiliation{\ECEPrinceton}
\author{Apoorv Jindal}\affiliation{\ECEPrinceton}
\author{Guangming Cheng}\affiliation{\IAC}
\author{Nan Yao}\affiliation{\IAC}
\author{Robert J. Cava}\affiliation{\ChemPrinceton}
\author{Andrew A. Houck}\affiliation{\ECEPrinceton}
\author{Nathalie P. de Leon}\affiliation{\ECEPrinceton}\email{npdeleon@princeton.edu}

\tableofcontents

\section{Transmission Electron Microscopy}
\label{section:B}

\begin{figure}[H]
    \centering
    \includegraphics[width=0.6\linewidth]{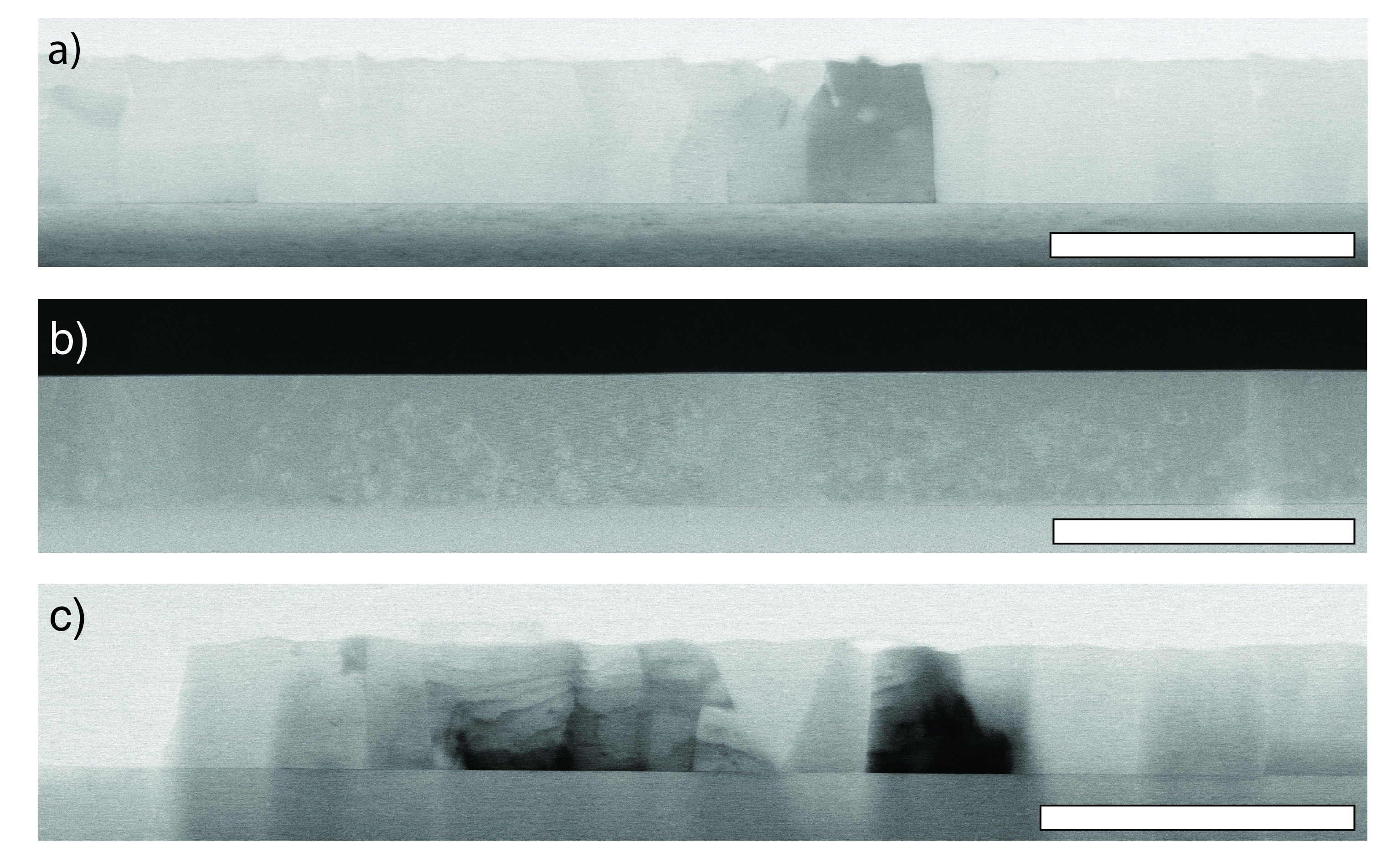}
    \caption{\textbf{Brightfield TEM of aluminum films.} Cross sections of aluminum films deposited at a) -72$\degree$ on c-Al$_2$O$_3$, b) 200$\degree$ on c-Al$_2$O$_3$ and c) 25$\degree$ on (100)-Si. The grain structure in a) and c) shows smaller and more irregular grains than in b), which shows larger and more uniform grains with a smoother film surface.}
    \label{fig:SI_CSTEM}
\end{figure}

\begin{figure}[H]
    \centering
    \includegraphics[width=0.6\linewidth]{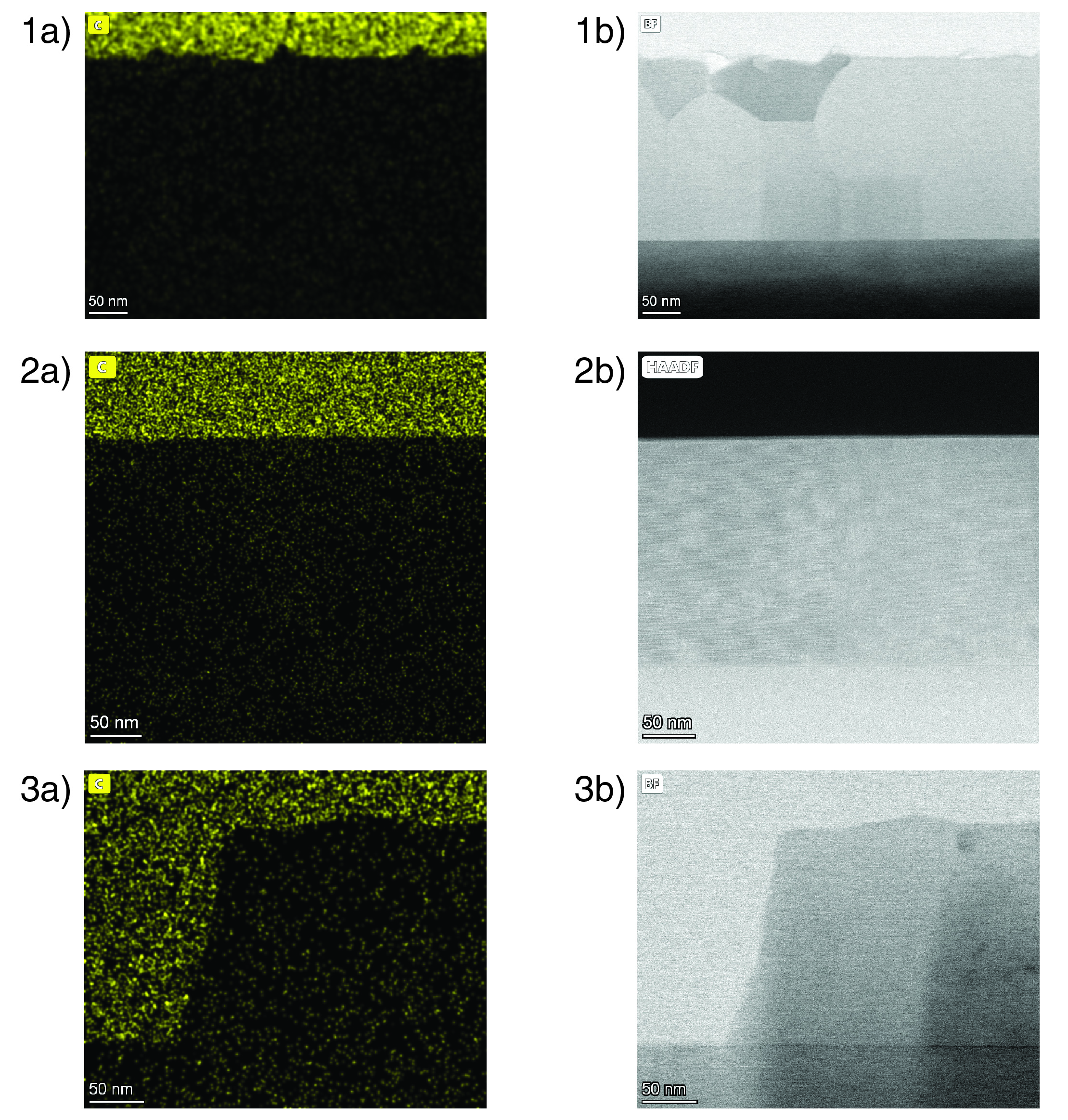}
    \caption{\textbf{EDS of aluminum-substrate interface.} Cross-sectional STEM-EDS of aluminum films deposited at 1) -72$\degree$ C on c-Al$_2$O$_3$, 2) 200$\degree$ C on c-Al$_2$O$_3$ and 3) 25$\degree$ C on (100)-Si. Each EDS scan (a) shows carbon signal which is absent at the interface, the location of which is corroborated using HAADF or brightfield STEM imaging of the same area (b). }
    \label{fig:SI_carbon}
\end{figure}

\begin{figure}[H]
    \centering
    \includegraphics[width=0.6\linewidth]{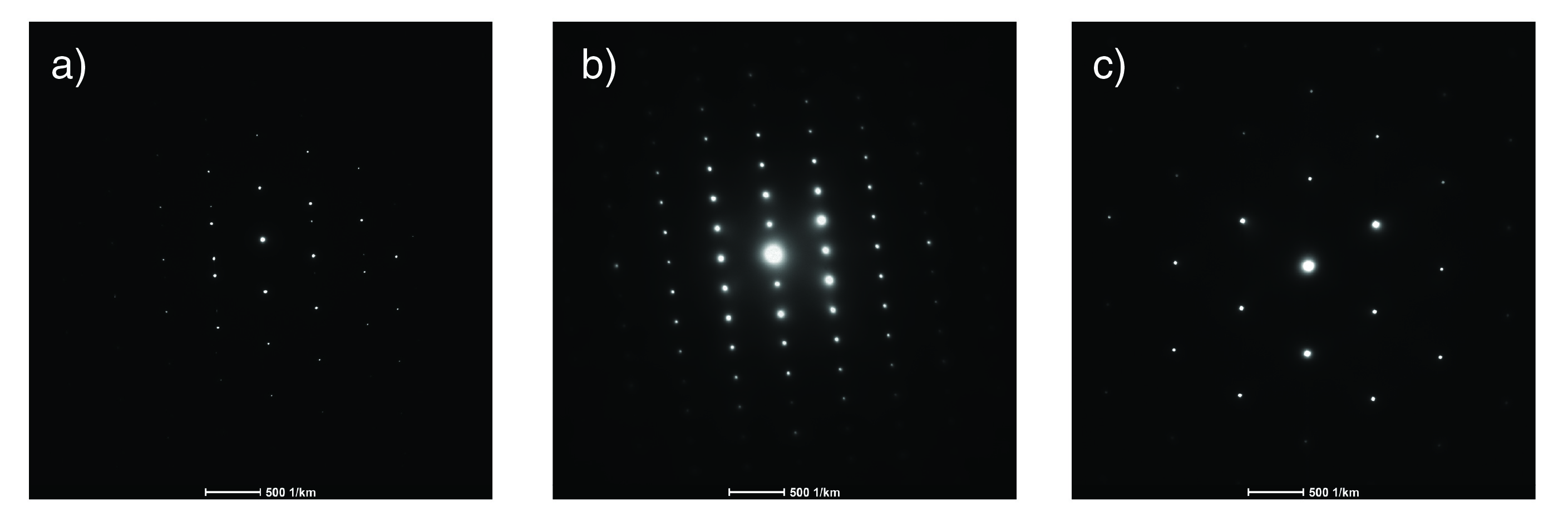}
    \caption{\textbf{Electron diffraction of aluminum films.} Cross-sectional TEM of aluminum films deposited at a) -72$\degree$ on c-Al$_2$O$_3$, b) 200$\degree$ on c-Al$_2$O$_3$ and c) 25$\degree$ on (100)-Si. All films exhibit cubic geometry.}
    \label{fig:SI_EDiff}
\end{figure}

We take scanning transmission electron microscope (STEM) images of three films deposited using different conditions to examine the cross-sectional grain sizes, as well as to examine the nature of the interface between aluminum and the substrate. First, examining the grains between films using brightfield STEM shows differences in the surface texture, grain structure and heteroepitaxy discussed in the main text. Mean grain sizes are calculated using Fig. \ref{fig:SI_CSTEM}, as AFM grain sizes were inaccurate due to the texture differences between grains and between films. 
We also perform energy dispersive x-ray spectroscopy (EDS) to examine carbon contamination at the metal-substrate interface. We do not detect carbon at the metal-substrate interface for any of the devices, including the device fabricated using liftoff (Fig. \ref{fig:SI_carbon}). We additionally show electron diffraction to verify the FCC cubic crystal structure of all three aluminum films (Fig. \ref{fig:SI_EDiff}).

\section{DC Resistivity and X-ray Diffraction}
\label{section:C}
\begin{figure}[H]
    \centering
    \includegraphics[width=\linewidth]{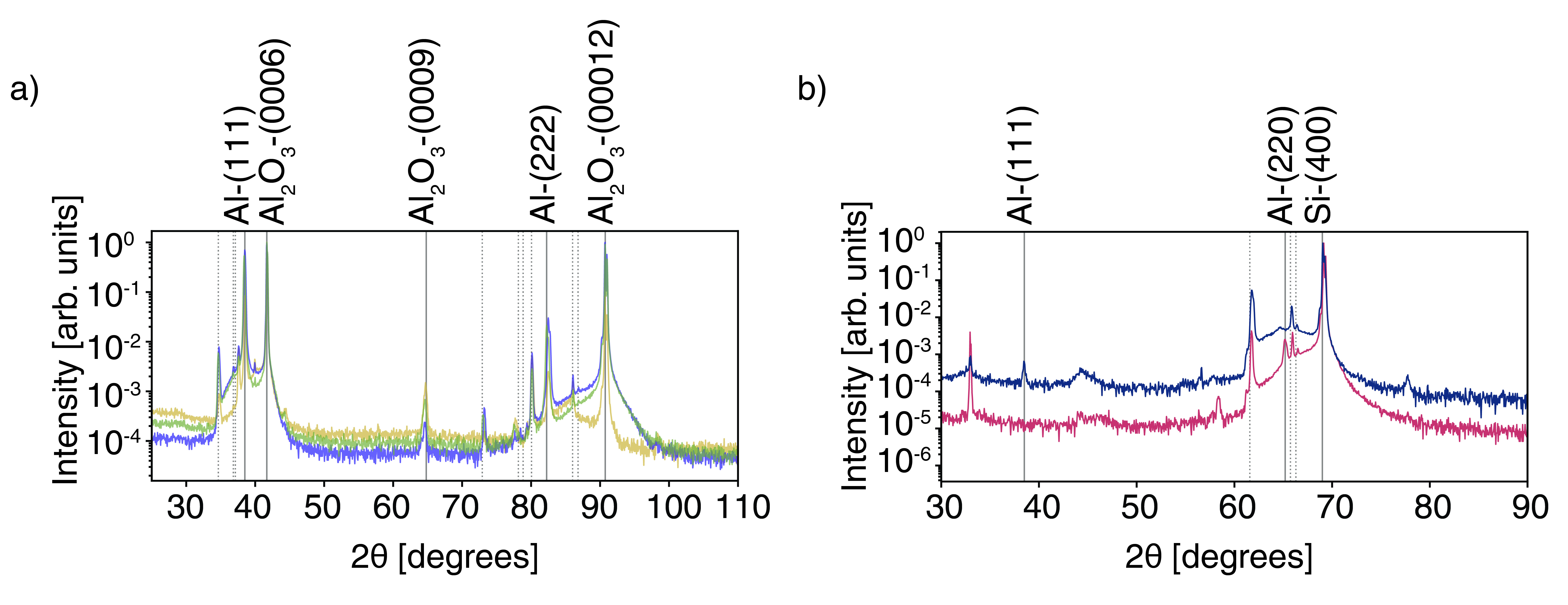}
    \caption{\textbf{XRD patterns for Al on c-Al$_2$O$_3$ and (100)-Si substrates.} Full spectra for 200 nm Al films deposited on a) c-Al$_2$O$_3$ and b) (100)-Si. All films deposited on c-Al$_2$O$_3$ show (111)-Al films with a secondary (222)-Al peak (purple: 200$\degree$ C, green: room temperature, yellow: -72$\degree$C). 
    The Al film deposited on (100)-Si (navy) at room temperature shows (111) orientation, while the 200$\degree$C film on (100)-Si (red) shows predominantly (220)-Al, which was not seen in any other deposited aluminum films. All dashed lines are from contaminant radiation due to tungsten deposition onto the x-ray tube. Unmarked peaks in the (100)-Si spectrum (bottom) are due to wafer miscut.}
    \label{fig:SI_XRD}
\end{figure}

We measure dc resistivity on all three films by four-point probe measurements using a Quantum Design PPMS Dynacool instrument with a He3 probe.
We measure x-ray diffraction on a Bruker D8 Advance instrument with a Copper K$\alpha$ source. Full spectra for 200 nm Al films deposited on c-Al$_2$O$_3$ (top) and (100)-Si are shown in Fig. \ref{fig:SI_XRD}. All films are primarily Al-(111) except the 200$\degree$C film deposited on (100)-Si, which is predominantly (220).

\section{Sidewall profiling}
\label{section:E} 
\begin{figure}[H]
    \centering
    \includegraphics[width=\linewidth]{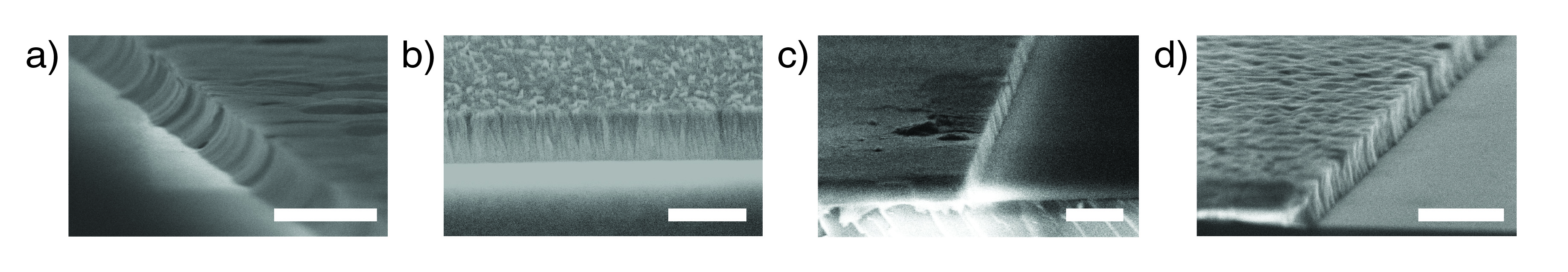}
    \caption{\textbf{Sidewall profiles of aluminum resonators.} a) Sidewall profile of 200 nm Al resonators deposited at 200$\degree$C etched with wet etch chemistry. The sidewall is concave into the metal. Scale bar is 400 nm. b) Sidewall profile of 200 nm Al resonators deposited at -72$\degree$C fabricated using liftoff. The sidewall is vertical and the top of the film has a rough texture. Scale bar is 400 nm. c) Sidewall profile of 200 nm Al resonators deposited at 200$\degree$C fabricated using reactive ion etch. The sidewall is vertical. Scale bar is 400 nm. d) Sidewall profile of 200 nm Al resonators deposited at 25$\degree$C fabricated using wet etch. The sidewall is vertical and the film has a lumpy texture. Scale bar is 400 nm. 
}
\label{fig:SI-sidewall}
\end{figure}

We note variation in the sidewalls of aluminum resonators with different deposition and etch conditions. In wet etched films (Fig. \ref{fig:SI-sidewall}a), a concave side  wall is seen for heated deposition films on sapphire and a straight edge on room temperature deposition films on Si (Fig. \ref{fig:SI-sidewall}d). In dry etched (Fig. \ref{fig:SI-sidewall}b) and liftoff devices (Fig.\ref{fig:SI-sidewall}c), we observe straight sidewalls. We do not observe strong correlations between performance and sidewall profile, but more measurements are needed to disentangle the contributions of fabrication techniques from sidewall geometry.

\section{Measurement Apparatus and Protocol}
\label{section:G}
We measure all devices in a Bluefors LD400 with a base mixing plate temperature of approximately 10-12 mK. The attenuation and filtering configuration as well as the  circle-fitting procedure for extracting $\mathrm{Q_{TLS,0}}$ has been previously reported in Ref. \cite{blandbahrami20252D}.

\section{Surface Participation Ratios}
\label{section:H} 
Similar to Refs. \cite{crowley2023disentangling} and \cite{blandbahrami20252D}, we use dc finite element simulation (HFSS) to find surface electric field participations for our CPW design. We assume a three nm dielectric with $\varepsilon = 10$ on all interfaces (metal-air, substrate-air, metal-substrate) and a 520 $\micro m $ substrate. We simulate assuming these thicknesses with various gaps between the center-pin and the ground plane in a 2D cross-section of a CPW resonator.

\section{X-ray photoelectron spectroscopy (XPS)}
\label{section:I}

We take all XPS data using a Thermo Fisher Nexsa G2 X-ray Photoelectron Spectrometer System using an Al K$\alpha$ = 1486 eV X-ray source. We take Al2p spectra in 0.05 eV steps, averaging over 10 scans with a spot size of 400 $\mu$m and a dwell time of 50 ms. We measure C1s and O1s spectra in 0.1 eV steps and average over 50 scans with a spot size of 400 $\mu$m and a dwell time of 50 ms. We shift C1s and O1s spectra in binding energy to C1s = 284.8 eV
and Al2p spectra to center the Al$^{0}$ peak at 72.6 eV \cite{ramirez2022testing}, as the adventitious carbon peak on aluminum can shift between 284 and 286 eV. We normalize each spectrum in intensity to the Al2p spectrum intensity. We fit each spectrum with a Shirley background from 70-80 eV and fit to a multicomponent model containing Lorentzian lineshapes for metallic peaks (Al$^{0}$) and Gaussian lineshapes for nonzero oxidation states (Al$\mathrm{_{int}}$,   Al$^{3+}$). Al$\mathrm{_{int}}$ encompasses a broad interface between metal and oxide containing aluminum suboxides which cannot be individually distinguished with our x-ray source \cite{ramirez2022testing}. We divide each Al2p peak into spin orbital components of Al2p$_{1/2}$ and Al2p$_{3/2}$ with a 2:1 intensity ratio and spin orbital splitting of 0.44 eV. The Al2p$_{3/2}$ peak centers were assigned using the oxidation state energies listed in Ref. \cite{ramirez2022testing} and constrained within 0.2 eV (0.5 for Al$^{3+}$). We calculate Al2p$_{1/2}$ parameters using the spin-orbital splitting parameters above.

After fitting, we calculate the area (intensity) of each component in the spectrum using fit parameters.  We use an area ratio of each component to find the approximate thickness of the total oxide including all non-zero oxidation states. We use area ratios of the metal to the oxide to calculate the total oxide thickness using Eqn. \ref{eqn:strohmeier} \cite{strohmeier1990esca}.

\begin{align}
\label{eqn:strohmeier}
 \mathrm{d = \lambda_{ox} \sin \theta \ln\left(\frac{N_{m}}{N_{ox}} \cdot \frac{I_{ox}}{I_{m}} \cdot \frac{\lambda_{m}}{\lambda_{ox}} +1 \right)}
\end{align}

In Eq. \ref{eqn:strohmeier}, $\ I_{m}$ ($\ I_{ox}$) is the intensity (peak area) of the metal (oxide), $\ N_{m}$ ($\ N_{ox}$) the volume density of metal atoms in the metal (oxide), and $\lambda_{m}$ ($\lambda_{ox}$) the inelastic mean free path of photoelectrons in the metal (oxide). $\theta = 90{\degree}$ is the angle of photoelectron ejection with respect to the surface. We take values for $\lambda_{m}$ and $\lambda_{ox}$ from Ref. \cite{shinosuka2015calculations}\cite{shinosuka2019calculations}\cite{powell2020practical} and $\ N_{m}$ and $\ N_{ox}$ from Ref. \cite{ramirez2022testing}.

\begin{figure*}
    \centering
    \includegraphics[width=0.5\linewidth]{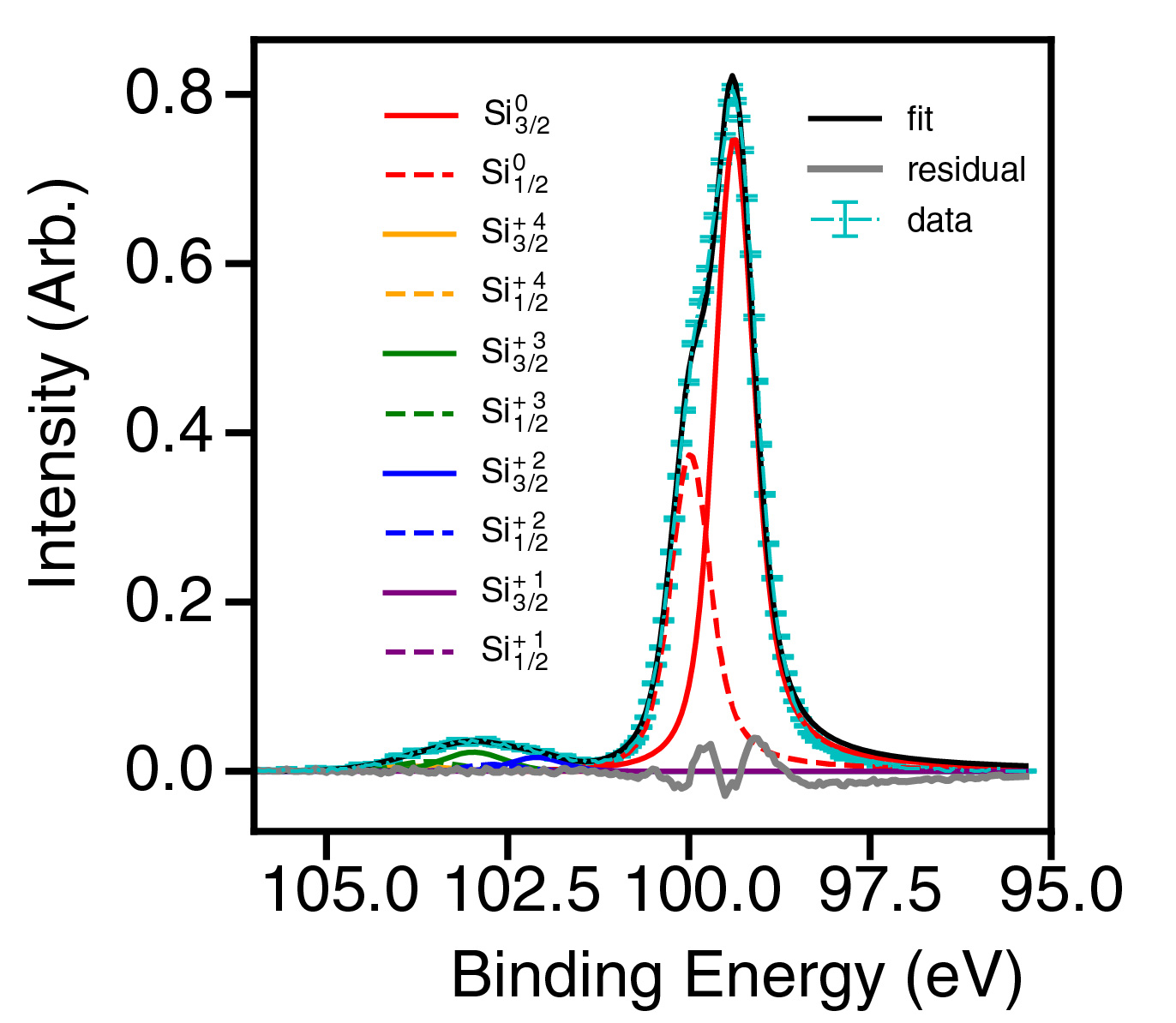}
    \caption{\textbf{Silicon surface characterization using X-ray Photoelectron Spectroscopy.}  Oxide thickness is calculated using the Strohmeier equation (Eqn. \ref{eqn:strohmeier}) and 5 oxidation states based on the fitting procedure in Ref. \cite{blandbahrami20252D}. }
    \label{fig:SISi}
\end{figure*}

To find the thickness of the SiO$_{x}$ left behind at the SA interface after 49\% HF treatment, we use the same fitting and literature values as in Ref. \cite{blandbahrami20252D} (Fig. \ref{fig:SISi}).

\noindent To measure the thickness of oxide or carbon residue on a time-sensitive device, such as the HF-treated devices that are cooled down within 2-4 hours, we take XPS measurements on a duplicate device when the dilution refrigerator has reached $\sim$ 1 $\times$ 10$^{-4}$ mTorr.  This ensures an accurate measurement of the surface of the device without slowing down the microwave measurement timeline.

\section{Loss modeling for aluminum resonators}
\label{section:J}
Dielectric losses in aluminum devices are dominated by surface losses. In this regime, we write dielectric losses in superconducting resonators as

\begin{align}
\label{equation1}
\mathrm{\frac{1}{Q_{TLS,0}}} = \mathrm{{p_{MS}}} \tan \delta_\mathrm{{MS}} + \mathrm{{p_{MA}}} \tan \delta_\mathrm{{MA}} + \mathrm{{p_{SA}}} \tan \delta_\mathrm{{SA}}
\end{align}

\noindent where t$_0$ = 3 nm is the simulated thickness of the metal-air (MA), substrate-air (SA), and metal-substrate (MS) interfaces, $\mathrm{{p_{MS}}}, \mathrm{{p_{MA}}},$ and $ \mathrm{{p_{SA}}}$ are electric field participation ratios found using HFSS simulation, and tan $\delta_\mathrm{{MS}}, $tan $\delta_\mathrm{{MA}},$ and tan $\delta_\mathrm{{SA}}$ are the loss tangents of the MS, MA and SA interfaces, respectively. 

\noindent To obtain the loss tangents in Fig. 3b in the main text, we recast the above equation in linear form.

\begin{equation}
\label{equation2}
\begin{aligned}
\mathrm{\frac{1}{Q_{TLS,0}}}
&= \mathrm{{p_{MS}}}\left(
\tan \delta_\mathrm{{MS}}
+ \mathrm{\frac{p_{MA}}{\mathrm{{p_{MS}}}}}\tan \delta_\mathrm{{MA}}
+ \mathrm{\frac{p_{SA}}{\mathrm{{p_{MS}}}}}\tan \delta_\mathrm{{SA}}
\right)\\
&= \mathrm{{p_{MS}}}\tan \delta
\end{aligned}
\end{equation}

\noindent where tan $\delta$ is the loss tangent extracted from the data in Fig. 3b in the main text. For each surface treatment, we therefore have

\begin{align}
\label{equation3}
 \tan \delta &= \tan \delta_\mathrm{{MS}} + \mathrm{\frac{p_{MA}}{\mathrm{{p_{MS}}}}} \tan \delta_\mathrm{{MA}} + \mathrm{\frac{p_{SA}}{\mathrm{{p_{MS}}}}} \tan \delta_\mathrm{{SA}}
 \end{align}

Using Eq. \ref{equation3}, we can write loss models for the HF-treated Al, HF-treated Al after 90 days, and untreated Al  surface loss tangents. For the MA interface oxide, we scale the loss term by the measured XPS thickness of each oxide, $\mathrm{t{_i}}$,  divided by the simulated thickness, $\mathrm{t_0}$. The extracted surface loss tangents for aluminum-on-sapphire and aluminum-on-silicon devices are the same; therefore we conclude that the MS and SA loss contributions are the same for the two substrates. We first compare the surface loss tangents from HF-treated Al (Eq. \ref{equation4}) and HF-treated Al after 90 days (Eq. \ref{equation5}) to find the materials-intrinsic loss tangent from amorphous $\mathrm{AlO_{x}}$:

\begin{align}
\label{equation4}
    \tan \delta_\mathrm{Al, HF} &= \mathrm{\frac{p_{MA}}{p_{MS}}}  \cdot  \mathrm{\frac{t_{HF, AlO_x}}{t_0}} \cdot \tan \delta_\mathrm{{AlO_x}} + \mathrm{\frac{p_{SA}}{p_{MS}}} \cdot \mathrm{\frac{t_{SA}}{t_0}} \cdot  \tan \mathrm{\delta_{SA} + \frac{t_{MS}}{t_0}} \tan \delta_\mathrm{MS}\\
    \label{equation5}
    \tan \delta_\mathrm{Al, HF 90 days} &= \mathrm{\frac{p_{MA}}{p_{MS}}}  \cdot  \mathrm{\frac{t_{HF 90 days, AlO_x}}{t_0}} \cdot \tan \delta_\mathrm{{AlO_x}} + \mathrm{\frac{p_{SA}}{p_{MS}}} \cdot \mathrm{\frac{t_{SA}}{t_0}} \cdot  \tan \mathrm{\delta_{SA} + \frac{t_{MS}}{t_0}} \tan \delta_\mathrm{MS}
\end{align}

\noindent Subtracting Eq. \ref{equation4} from Eq. \ref{equation6} and solving for $\tan \delta_\mathrm{{AlO_x}}$, we find:

\begin{equation}
\label{equation6}
\begin{aligned}
\tan \delta_\mathrm{{AlO_x}} &=
\mathrm{\frac{p_{MS}}{p_{MA}}}
\left(\tan \delta_\mathrm{Al, HF 90 days} - \tan \delta_\mathrm{Al, HF}\right)
\cdot
\mathrm{\frac{t_0}{t_{HF 90 days, AlO_x} - t_{HF, AlO_x}}}\\
&= (1.74 \pm 0.7) \times 10^{-2}
\end{aligned}
\end{equation}

Next, we use $\tan \delta_\mathrm{{AlO_x}}$ to find the combined loss from the MS and MA interfaces:

\begin{equation}
\label{equation7}
\begin{aligned}
\mathrm{\frac{p_{SA}}{p_{MS}}}\cdot \mathrm{\frac{t_{SA}}{t_0}}\cdot \tan \delta_\mathrm{SA}
+ \mathrm{\frac{t_{MS}}{t_0}}\tan \delta_\mathrm{MS}
&= \tan \delta_\mathrm{Al, HF}
- \tan \delta_\mathrm{{AlO_x}}\cdot \mathrm{\frac{p_{MA}}{p_{MS}}}\cdot \mathrm{\frac{t_{HF, AlO_x}}{t_0}}\\
&= (6.19 \pm 4.96) \times 10^{-4}
\end{aligned}
\end{equation}

After isolating both the loss from the MA interface when solely made up of $\mathrm{AlO_{x}}$ and the MA and MS combined, we now turn our attention to the untreated Al surface loss tangent, which includes loss due to fabrication residue in addition to the aforementioned sources of loss. We measure surface HC on the MA interface using XPS, normalized in intensity. The thickness of the hydrocarbon layer ($\mathrm{t_{i, HC}}$) is found by converting signal to number of carbon monolayers, assuming each monolayer is $\sim$ 0.5 nm thick and each 7.6\% atomic percent of carbon is one monolayer \cite{palinkas2022composition}\cite{sangtawesin2019origins}. We find that the carbon layer thickness is roughly 0.52 nm on the surface of the untreated aluminum device. We assume fabrication residue of this magnitude covers the SA and MA interfaces. Using these assumptions, we can calculate the fabrication hydrocarbon loss tangent by solving the expression for total loss in untreated devices, Eq. \ref{equation8}, for $\mathrm{\tan\delta_{HC}}$ (Eq. \ref{equation9}).

\begin{align}
\label{equation8}
\tan \delta_\mathrm{Al, untreated} &= \mathrm{\frac{p_{MA}}{p_{MS}}}  \cdot  \mathrm{\frac{t_{untreated, AlO_x}}{t_0}} \cdot \tan \delta_\mathrm{{AlO_x}} +  \mathrm{\frac{p_{MA} + p_{SA}}{p_{MS}} } \cdot \mathrm{\frac{t_{i, HC}}{t_0}} \tan \delta_\mathrm{HC} + \mathrm{\frac{p_{SA}}{p_{MS}}} \cdot  \mathrm{\frac{t_{SA}}{t_0}} \cdot \tan \mathrm{\delta_{SA} + \frac{t_{MS}}{t_0}} \tan \delta_\mathrm{MS}
\end{align}

\begin{equation}
\label{equation9}
\begin{aligned}
\tan \delta_\mathrm{HC} &=
\mathrm{\frac{t_0}{t_{i, HC}}}\,\mathrm{\frac{p_{MS}}{p_{MA} + p_{SA}}}\cdot \\
& \ \ \Bigg(
\tan \delta_\mathrm{Al, untreated}
- \mathrm{\frac{p_{MA}}{p_{MS}}}\cdot \mathrm{\frac{t_{untreated, AlO_x}}{t_0}}\cdot \tan \delta_\mathrm{{AlO_x}} - \mathrm{\frac{p_{SA}}{p_{MS}}}\cdot \mathrm{\frac{t_{SA}}{t_0}}\cdot \tan \delta_\mathrm{SA}
- \mathrm{\frac{t_{MS}}{t_0}}\tan \delta_\mathrm{MS}
\Bigg)\\
&= (3.89 \pm 1.12) \times 10^{-3}
\end{aligned}
\end{equation}

We then compare the fraction of TLS loss arising from each material interface for untreated aluminum by multiplying each loss tangent by the SPR for the region it has been modeled on, and then dividing by the loss tangent for untreated aluminum. We find that aluminum oxide accounts for $ (52.4 \pm 21.5)\%$, hydrocarbons for $(27.7 \pm 7.2) \%$, and the MS and SA surfaces combined for $(19.8 \pm 15.9) \%$ of surface TLS losses, making aluminum oxide the dominant source of linear absorption loss in aluminum resonators. 

To find the materials intrinsic loss tangents of $\mathrm{AlO_x}$ and the MS interface that best represent transmon qubit operating parameters, we rescale each fitted $\mathrm{Q_{TLS,0}}$ using \ref{equation10} to find $\mathrm{Q_{TLS}(\bar n = 1)}$ at base temperature (10 mK) \cite{crowley2023disentangling}.

\begin{align}
\label{equation10}
\mathrm{Q_{TLS}(\bar n, T)} &= \mathrm{Q_{TLS,0}} \frac{\sqrt{1+(\frac{\bar n^{\beta_2}}{D T^{\beta_1}}) \tanh{(\frac{\hbar \omega }{2 k_B T})}}}{\tanh{(\frac{\hbar \omega }{2 k_B T})}}
\end{align}

\begin{figure*}
    \centering
    \includegraphics[width=0.5\linewidth]{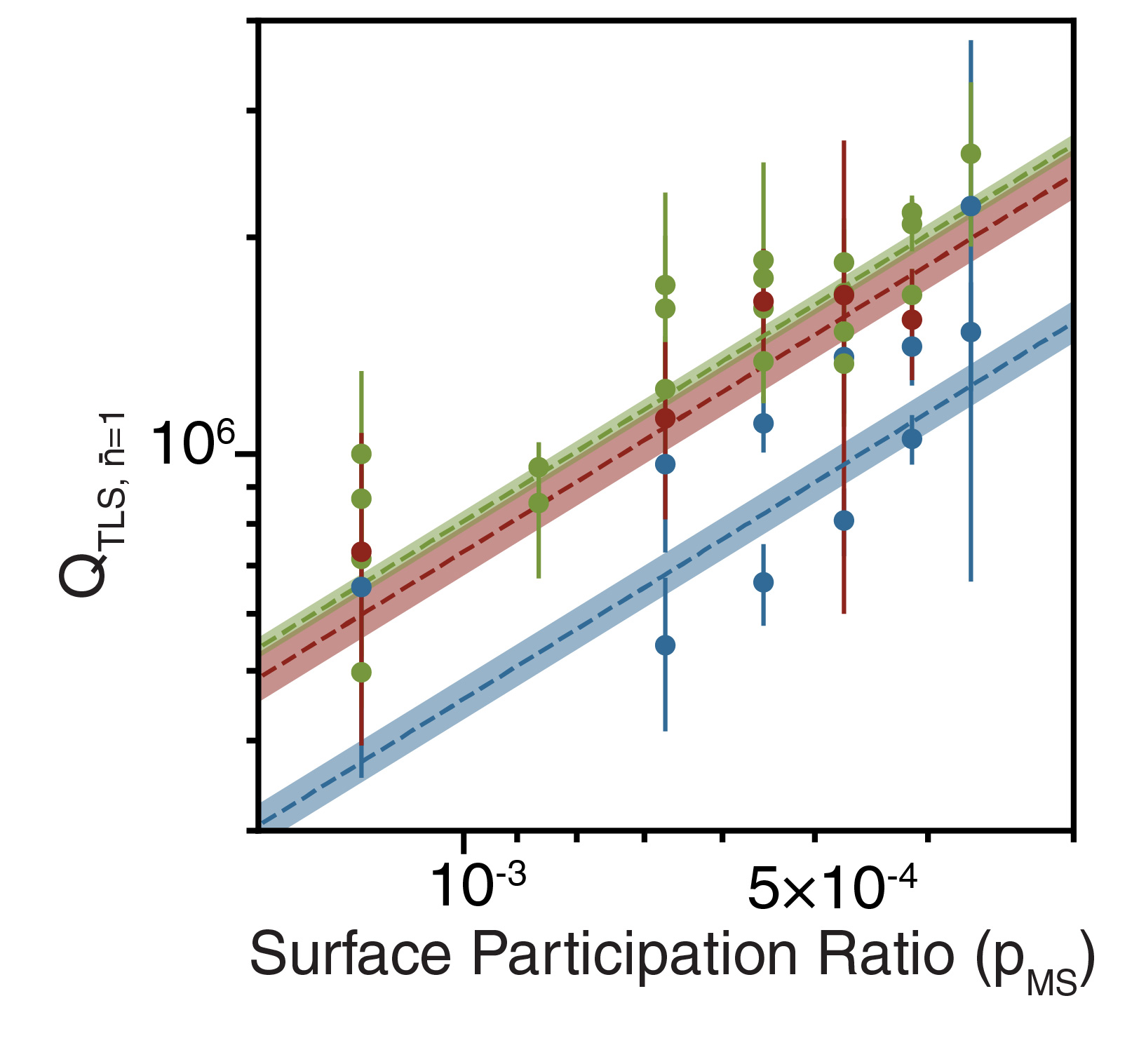}
    \caption{\textbf{Dependence of $\mathbf{\mathrm{Q_{TLS}(\bar n = 1)}}$ on surface participation.} We find that the linear dependence of $\mathrm{Q_{TLS,0}}$ with surface participation ratio is preserved when transforming each point to $\mathrm{Q_{TLS}(\bar n = 1)}$ using T=10 mK. The three surface conditions used in the model are shown: Al treated with HF and measured after 2 hours (green), Al treated with HF and measured after 90 days (red), and untreated Al (blue). We find that $\mathrm{\tan \delta_{HF,\ \bar n = 1} = (12.39 \pm 0.4) \times 10^{-4}}$, $\mathrm{\tan \delta_{HF \ 90 \ days,\ \bar n = 1} = (13.66 \pm 1.0) \times 10^{-4}}$,  and $\mathrm{\tan \delta_{untreated,\ \bar n = 1} = (21.8 \pm 1.4) \times 10^{-4}}$.}
    \label{fig:SI_QTLSn=1}
\end{figure*}

\noindent We then fit surface loss tangents to the surface conditions that are used in the above loss model (Fig. \ref{fig:SI_QTLSn=1}).

Performing the same analysis as for the case of linear absorption, we find $\mathrm{\tan \delta_{AlO_x,\ \bar n = 1} = (2.99 \pm 0.23) \times 10^{-3}}$ and $\mathrm{\tan \delta_{MS \ and \ SA,\ \bar n = 1} = (1.04 \pm 0.01) \times 10^{-3}}$. These values vary significantly from their linear absorption counterparts, as $\mathrm{\tan \delta_{AlO_x,\ \bar n = 1}}$ is over 5 times less than $\mathrm{\tan \delta_{AlO_x}}$. We will use the loss tangents valid in the single photon regime for the loss budgeting of Ta-on-Si transmon qubits in section \ref{section:K}.

\section{Loss modeling for aluminum Josephson junction leads}
\label{section:K}

We use HFSS simulation combined with experimental data to find the approximate limiting Q from just the leads of an Al/$\mathrm{AlO_{x}}$/Al Josephson junction. The rapidly varying fields around the junction tunnel barrier make this region difficult to simulate accurately, so we exclude the sandwiched junction dielectric, as well as a small region around the barrier \cite{wang2015surface}. To best model this system, we express the total surface loss of a qubit to be the sum of the dominant surface losses of its parts (Eq. \ref{equation11}). For the surface losses from the capacitor pads, we expect a similar SPR distribution of $\mathrm{p_{MS}}$, $\mathrm{p_{SA}}$, and $\mathrm{p_{MA}}$ as in the lumped element (LE) resonator geometry, so we use the loss tangent of Ta-on-Si resonators at $\bar n = 1$ multiplied by the participation of the capacitor pad surfaces to represent total TLS loss contributions from the capacitor pads. For the Josephson junction leads, we instead use the individual materials loss tangents for $\mathrm{AlO_{x}}$ and the combined SA and MS interfaces found in the main text, as we expect the electric field distribution in this region to be different than in a LE geometry. 

\begin{align}
\label{equation11}
\mathrm{\frac{1}{Q_{70 \mu m}}} &= \mathrm{p_{capacitor}}\tan \delta_\mathrm{device, capacitor} + \mathrm{p_{MA,junction\ leads}}\tan\delta_\mathrm{MA, junction\ leads} + \mathrm{p_{MS,junction \ leads}}\tan\delta_\mathrm{MS, junction \ leads}
\end{align}

Next, we simulate a simplified 3D qubit structure in HFSS and find $\mathrm{p_{capacitor}}$, $\mathrm{p_{MA,junction \ leads}}$, and $\mathrm{p_{MS,junction \ leads}}$. We assume that these three surfaces are the dominant source of surface loss in 2D qubit geometries. 

We then write an equation relating the surface losses of each circuit component to the limiting qubit Q, assuming only surface losses. From HFSS simulation, we find that $\mathrm{p_{capacitor}} = \mathrm{p_{MS, Ta}} = 0.983 \times 10^{-4}$, the surface participation ratio of the interfaces of the Ta capacitor pads, $\mathrm{p_{MS, junction \ leads}} = \mathrm{p_{MS, Al \ leads}} = 0.160 \times 10^{-4}$, the surface participation ratio of the MS interfaces of the Al Josephson junction leads, and $\mathrm{p_{MA, junction \ leads}} = \mathrm{p_{MA, Al \ leads}} = 0.013 \times 10^{-4}$, the SPR from the MA interface of the Al Josephson junction leads. These surface participations correspond to a qubit on silicon assuming a 70 $\mu$m gap and 100 nm etching into the silicon substrate between capacitor pads, which aligns the most accurately with \cite{blandbahrami20252D}. For the surface loss tangents for specific devices or materials, we have two choices: using the loss due to the linear absorption of TLSs, found by finding surface loss tangents as in Eq. \ref{equation2}, or by finding surface loss tangents for the expected steady state transmon operating conditions $\ \bar n=1$. To do this, we rescale each $\mathrm{Q_{TLS,0}}$ using Eq. \ref{equation10} and then calculate the same surface loss tangents (see section \ref{section:I}).

First we will do the procedure assuming losses from linear absorption. For the capacitor pads, we use $\tan \delta_\mathrm{Ta-on-Si} = (11.3 \pm 0.5) \times 10^{-4}$ from Ref. \cite{blandbahrami20252D}. For the Josephson junction leads, we use the loss tangent $\tan \delta_\mathrm{AlO_x} = (1.74 \pm 0.7) \times 10^{-2}$ to scale the MA interface and $\tan \delta_\mathrm{MS} = (6.19 \pm 4.96) \times 10^{-4}$, the sum of the MS and the SA for aluminum on sapphire/silicon in the main text. We note that the values in the main text for the loss tangent of MS and SA include the surface participation ratios for the resonator geometry, so this calculation will produce an underestimation for the minimum Q from solely junction losses. Using these values, we evaluate Eq. \ref{equation12}.

\begin{align}
\label{equation12}
\mathrm{\frac{1}{Q_{70 \mu m}}} &= \mathrm{p_{MS, capacitor}}\tan \delta_\mathrm{Ta-on-Si} + \mathrm{p_{MA,junction \ leads}}\tan\delta_\mathrm{AlO_{x}} + \mathrm{p_{MS,junction \ leads}}\tan\delta_\mathrm{Al, MS} \\
&= 0.983 \times 10^{-4} \times (11.3 \pm 0.5) \times 10^{-4} + 0.013 \times 10^{-4} \times (1.74 \pm 0.7) \times 10^{-2}\\ 
& \ \ \ \ + 0.160 \times 10^{-4} \times (6.19 \pm 4.96) \times 10^{-4}  \\
& = (1.436 \pm 0.130) \times 10^{-7} \\
& = \frac{1}{(7.0 \pm 0.4) \times 10^6}
\end{align}

For  the case of $\bar n = 1$, we solve the loss model in section \ref{section:J} again using the $\bar n$ surface loss tangents and find $\tan \delta_\mathrm{AlO_x,\ \bar n = 1} = (2.99 \pm 0.23) \times 10^{-3}$ and $\tan \delta_\mathrm{MS, \ \bar n = 1} = (10.4 \pm 0.1) \times 10^{-4}$. For the capacitor pads, we use $\tan \delta_\mathrm{Ta-on-Si} = (7.8 \pm 0.4) \times 10^{-4}$ extrapolated from Ref. \cite{blandbahrami20252D} using Eq. \ref{equation10}.

\begin{align}
\label{equation13}
\mathrm{\frac{1}{Q_{70 \mu m, \bar n=1}}} &= \mathrm{p_{MS, capacitor}}\tan \delta_\mathrm{Ta-on-Si, \ \bar n=1} + \mathrm{p_{MA,junction \ leads }}\tan\delta_\mathrm{AlO_{x}, \ \bar n=1} + \mathrm{p_{MS,junction \ leads}}\tan\delta_\mathrm{Al, MS, \ \bar n=1} \\
&= 0.983 \times 10^{-4} \times (7.8 \pm 0.4) \times 10^{-4} + 0.013 \times 10^{-4} \times (2.99 \pm 0.23) \times 10^{-3} \\
& \ \ \ \ + 0.160 \times 10^{-4} \times (10.4 \pm 0.1) \times 10^{-4}  \\
& = (0.972 \pm 0.039) \times 10^{-7} \\
& = \frac{1}{(10.3 \pm 0.4) \times 10^6}
\end{align}

When comparing these different methods, we find that the average qubit Q using linear absorption loss tangents is $(7.0 \pm 0.4) \times 10^{6}$, while the average qubit Q using steady state qubit operating condition loss tangents is  ($(10.2 \pm 0.4) \times 10^{6}$). The average Q value using $\bar n = 1$ loss tangents agrees with the average measured Q for Ta-on-Si qubits in Ref. \cite{blandbahrami20252D}, $\mathrm{Q_{avg}} = (9.74 \pm 0.33) \times 10^6 $. We can also compare the two sources of loss ($\bar n =1$); in the current state of the art, $(21.1 \pm 0.9)$\% of TLS losses can be expected from the amorphous surfaces in the Josephson junction leads, and $(78.8 \pm 0.9)$\% from the amorphous surfaces on the capacitor pads. These numbers are found by dividing $\mathrm{p_{MS, capacitor}}\tan \delta_\mathrm{Ta-on-Si,\ \bar n=1}$ by the total RHS for the fraction of loss from the Ta capacitor pads, and vice versa for losses from the Al junction leads.

\begin{align}
\label{equation14}
 \mathrm{TLS, Cap} 
 & = \frac{\mathrm{p_{MS, capacitor}}\tan \delta_\mathrm{Ta-on-Si,\ \bar n=1} \times 100}{\mathrm{p_{MS, capacitor}}\tan \delta_\mathrm{Ta-on-Si,\ \bar n=1} + \mathrm{p_{MA,junction \ leads}}\tan\delta_\mathrm{AlO_x, \ \bar n=1} + \mathrm{p_{MS,junction \ leads}}\tan\delta_\mathrm{Al, MS, \ \bar n=1}}  \\
& = (78.8 \pm 0.9)\%
\end{align}

\begin{align}
\label{equation15}
\mathrm{TLS, JJ} &=
 \frac{\mathrm{p_{MA,junction \ leads}}\tan\delta_\mathrm{AlO_{x}, \ \bar n=1} + \mathrm{p_{MS,junction \ leads}}\tan\delta_\mathrm{Al, MS, \ \bar n=1}}{\mathrm{p_{MS, capacitor}}\tan \delta_\mathrm{Ta-on-Si,\ \bar n=1} + \mathrm{p_{MA,junction \ leads}}\tan\delta_\mathrm{AlO_{x}, \ \bar n=1} + \mathrm{p_{MS,junction \ leads}}\tan\delta_\mathrm{Al, MS, \ \bar n=1} } \times 100 \\
 & = (21.1 \pm 0.9)\% \\
\end{align}

Though we can not accurately simulate electric field participation in the junction barrier oxide, we can estimate the rough internal energy fraction between the Josephson junction and the capacitor pads. To do this, we model the Josephson junction as a parallel plate capacitor with an area of $\mathrm{(200 nm \pm 50 nm) \times (200 nm \pm 50 nm)}$ , a dielectric thickness of (2 $\pm$ 0.5) nm, and an estimated relative permittivity of $\mathrm{ \varepsilon_r} = 9$. We find the capacitance to be $\mathrm{C_{JJ} = (1.59 \pm 0.89 ) \ fF}$. We compare this to the simulated capacitance of the transmon design used in \cite{blandbahrami20252D}, $\mathrm{C_{shunt}}$ = 96 fF. We find that the percentage of energy in the Josephson junction, $\mathrm{\frac{C_J}{C_J+C_{shunt}}}$, equals $(1.6 \pm 0.9)$\%. We can write Eq. \ref{equation16} to express the sum of expected sources of TLS losses in the Ta-on-Si transmon qubits from Ref. \cite{blandbahrami20252D}.

\begin{align}
\label{equation16}
\mathrm{\frac{1}{Q_{70 \mu m, \bar n=1}}} &= \mathrm{\frac{C_{shunt}}{C_{shunt}+C_{JJ \ barrier}}}(\mathrm{p_{MS, capacitor}}\tan \delta_\mathrm{Ta-on-Si, \ \bar n=1} + \mathrm{p_{MA,junction \ leads }}\tan\delta_\mathrm{AlO_{x}, \ \bar n=1}\\ 
& +\mathrm{p_{MS,junction \ leads}}\tan\delta_\mathrm{Al, MS, \ \bar n=1}) + \mathrm{\frac{C_{JJ \ barrier}}{C_{shunt}+C_{JJ \ barrier}}}\left( \tan\delta_\mathrm{AlO_{x} barrier, \ \bar n=1}\right)\ \\
\end{align}

Using the measured average qubit $\mathrm{Q_{70 \mu m, \bar n=1}} = (9.74 \pm 0.33) \times 10^6$ from Ref. \cite{blandbahrami20252D}, we can solve for $\mathrm{\tan\delta_\mathrm{AlO_{x} \ barrier, \ \bar n=1}}$. We find $\mathrm{\tan\delta_\mathrm{AlO_{x} \ barrier, \ \bar n=1}} = (4.3 \pm 3.7)\times 10^{-7}$, which is roughly three orders of magnitude lower than that of $\tan\delta_\mathrm{AlO_x, \bar n = 1}$. Others in literature have also found similar numbers on the order of $10^-7$ to $10^-8$ \cite{mamin2021merged} \cite{kim2011decoupling}. Therefore, an estimate for the dielectric losses from the junction barrier like that used in Ref.\cite{blandbahrami20252D}  would be 
$\\ \mathrm{\tan \delta_{JJ \ barrier} = (7.01 \pm 5.29) \times 10^{-9}}$, corresponding to a limiting Q of 142 million and a contribution of 6.8\% of the total loss budget in Ta on Si transmon qubits (Al JJ leads = 19.7\%, Ta capacitor pads = 73.5\%).

Our interpretation of this unexpectedly low loss tangent for amorphous $\mathrm{AlO_x}$ in the junction barrier follows from the lowest values in literature for the spectral density of TLS per area for Josephson junctions like that in Ref. \cite{blandbahrami20252D} ($\mathrm{A_{JJ, dielectric}}$ = 200  nm $\times$ 200 nm = 4 $\mathrm{\times 10^{-2} \micro m^2, V_{JJ, dielectric}}$ = $\mathrm{A_{JJ, dielectric} \times}$ 2  nm = 8 $\mathrm{\times 10^{-5} \mu m^3}$). In junctions with similar and larger areas, values of $\sim$ 0.5-1.5 TLS per GHz per $\micro m^2$  and 250-760 TLS per GHz per $\micro m^3$ have been reported \cite{osman2023mitigation, martinis2005decoherence, kline2009josephson, lisenfeld2019electric, bilmes2022probing}. This corresponds to a predicted 0.02 to 0.06 TLS per GHz for a 200 nm by 200 nm area junction, signifying that it is not only reasonable but expected for a TLS to be more often absent than not in an $\mathrm{AlO_x}$ junction barrier, especially given the ultra-high vacuum deposition employed in Ref. \cite{blandbahrami20252D}. In the extremely common absence of this TLS, we believe this loss tangent represents the loss tangent of sapphire, which is measured in literature as $<\ 5 \times 10^{-7}$ \cite{read2023precision}. 

\bibliographystyle{unsrtnat}
\bibliography{biblio}